\documentclass[12pt,hidelinks]{article}
\usepackage[latin9]{inputenc}
\usepackage{geometry}
\usepackage{array}
\usepackage{varwidth}
\usepackage{tikz}
\usepackage{amsmath}
\usepackage{amssymb}
\usepackage{graphicx}
\usepackage{setspace}
\usepackage{esint}
\usepackage[authoryear]{natbib}
\usepackage[]
 {hyperref}

\makeatletter

\providecommand{\tabularnewline}{\\}
\newenvironment{cellvarwidth}[1][t]
    {\begin{varwidth}[#1]{\linewidth}}
    {\@finalstrut\@arstrutbox\end{varwidth}}
\providecolor{lyxadded}{rgb}{0,0,1}
\providecolor{lyxdeleted}{rgb}{1,0,0}
\usetikzlibrary{calc}
\newcommand{\lyxobjectsout}[1]{%
  \bgroup%
  \color{lyxdeleted}%
  \tikz{
    \node[inner sep=0pt,outer sep=0pt](lyxdelobj){#1};
    \draw($(lyxdelobj.south west)+(2em,.5em)$)--($(lyxdelobj.north east)-(2em,.5em)$);
  }
  \egroup%
}

\DeclareRobustCommand{\lyxdisplayobjdeleted}[4][]{%
  \ifx#4\empty\else%
     \texorpdfstring{\leavevmode\\\lyxobjectsout{\parbox{\linewidth}{#4}}}{}%
  \fi%
}
\DeclareRobustCommand{\lyxudisplayobjdeleted}[4][]{%
  \ifx#4\empty\else%
     \texorpdfstring{\leavevmode\\\raisebox{-\belowdisplayshortskip}{%
                \lyxobjectsout{\parbox[b]{\linewidth}{#4}}}}{}%
     \leavevmode\\%
  \fi%
}

\usepackage{tikz}
\usepackage{kpfonts}  
\usetikzlibrary{arrows,shapes}
\usetikzlibrary{decorations.pathreplacing,calc}
\usepackage{tabularx}
\usepackage{bbm}
\usepackage{pdfpages}
\usepackage{amsfonts}
\usepackage{ctable}
\usepackage{rotating}
\usepackage{pdflscape}
\usepackage{url}
\usepackage{graphicx,xcolor}

\usepackage{amsmath}
\usepackage[para]{threeparttable}

\usepackage{units}

\usepackage{hyperref}
\hypersetup{colorlinks,linkcolor={blue},citecolor={blue},urlcolor={red}}

\newcounter{appendix}

\date{}

\makeatother

\begin{document}
\begin{titlepage} 

\title{\singlespacing{}{\Large\textbf{Demand Estimation with Text and Image Data\thanks{We thank Steve Berry, Alex Burnap, Matt Gentzkow, Joonhwi Joo, Philip
Haile, Yufeng Huang, Jon McClure, Carl Mela, Sanjog Misra, Olivia
Natan, Matt Osborne, Katja Seim, Adam Smith, Yuyan Wang, Kevin Williams
and seminar participants at Bocconi University, Chicago Booth, London
Business School, Stockholm School of Economics, University of Cologne,
Nova SBE, Marketing Science Conference 2022, QME Conference 2025,
SICS Conference 2024, AI/ML Yale Conference, and SED Conference for
helpful comments. We also thank Celina Park, Tanner Parsons, James
Ryan, Janani Sekar, Andrew Sharng, Franklin She, Adam Sy, and Vitalii
Tubdenov for outstanding research assistance. Contact information:
Giovanni Compiani (gio1compiani@gmail.com), Ilya Morozov (ilya.morozov@kellogg.northwestern.edu),
Stephan Seiler (stephan.a.seiler@gmail.com). Giovanni Compiani gratefully
acknowledges support from the Center for Applied AI at Booth.}}}}
\maketitle
\begin{center}
\begin{tabular}{ccc}
\textbf{Giovanni Compiani} & \textbf{Ilya Morozov} & \textbf{Stephan Seiler}\tabularnewline
University of Chicago & Northwestern University & Imperial College London\tabularnewline
 &  & \& CEPR\tabularnewline
\end{tabular}
\par\end{center}

\begin{center}
{\LARGE\vspace{0.1in}
}{\LARGE\par}
\par\end{center}

We propose a demand estimation approach that leverages unstructured
data to infer substitution patterns. Using pre-trained deep learning
models, we extract embeddings from product images and textual descriptions
and incorporate them into a mixed logit demand model. This approach
enables demand estimation even when researchers lack data on product
attributes or when consumers value hard-to-quantify attributes such
as visual design. Using a choice experiment, we show this approach
substantially outperforms standard attribute-based models at counterfactual
predictions of second choices.  We also apply it to 40 product categories
offered on Amazon.com and consistently find that unstructured data
are informative about substitution patterns.

{\small\vspace{0.1in}
}{\small\par}

\begin{singlespace}
\noindent{\small{}{\small\textbf{Keywords:}}{\small{} Demand Estimation,
Unstructured Data, Deep Learning}}{\small\par}

\noindent{\small{} }{\small\par}
\end{singlespace}

{\LARGE\vfill{}
}{\LARGE\par}

\end{titlepage}

\begin{onehalfspace}
\pagebreak{}
\end{onehalfspace}

\section{Introduction}

Many problems in economics and marketing---such as merger analysis,
tariff evaluation, and optimal pricing---require estimating demand
for differentiated products. A standard approach has been to estimate
demand models that capture substitution through the similarity of
product attributes. While common, this approach faces two practical
challenges. First, researchers rarely observe all choice-relevant
attributes. Instead, they rely on third-party data in which attributes
are chosen based on unknown criteria, or gather their own data, subjectively
choosing which attributes to collect.  Second, consumers often consider
visual design and functional benefits of products, dimensions that
are difficult to capture through observed attributes.\footnote{This criticism of attribute-based models has a long history in the
economics literature. For example, \citet[p.229]{hausman1994valuation}
skeptically remarks that applying such models to French champagne
choices would require researchers to somehow quantify the bubble content.}

In this paper, we show how researchers can incorporate unstructured
data---such as product images, descriptions, or reviews---in demand
estimation to recover substitution patterns.  Using pre-trained deep
learning models, we transform images and texts into vector representations
(\textsl{embeddings}) and apply Principal Component Analysis (PCA)
to reduce dimensionality and capture the main dimensions of product
differentiation. We incorporate these principal components into a
logit model with random coefficients, treating them as researchers
usually treat observed product attributes. Lastly, we implement a
model selection algorithm to choose the type of unstructured data
and deep learning model that extract the strongest signal  of substitution.

This approach lets researchers incorporate hard-to-quantify product
attributes, such as visual design in images and functional benefits
from text, while avoiding subjective choices about which attributes
to collect. It relies on product images and textual descriptions,
which are widely available even in markets where collecting product
attributes is challenging. Given the prevalence of unstructured data,
the approach offers a valuable addition to the toolbox of empirical
researchers. To ease adoption, we provide a Python package, \textit{DeepLogit},
alongside this paper.\footnote{The package is available on PyPI and in the public GitHub repository:
\href{https://github.com/deep-logit-demand/deeplogit}{github.com/deep-logit-demand/deeplogit}.
Our package heavily borrows from the existing package \textit{Xlogit}
for GPU-accelerated estimation of mixed logit models, developed by
\citet{arteaga2022xlogit}.}

We validate this approach empirically. A key challenge is that substitution
patterns are unobserved, so there is no ground truth to test model
predictions. An ideal dataset would let us estimate demand precisely
from consumers' choices and then evaluate how well the model predicts
an empirical measure of substitution not used in estimation. To achieve
this, we design an experiment in which participants choose a book
from a list of options, and we elicit both first and second choices.
We randomize prices and rankings of all options. This variation lets
us estimate substitution patterns using only first choices and without
instruments, while reserving second-choice data for validation. We
then test how well the model predicts second choices counterfactually.\footnote{We refer to second choice predictions as ``counterfactual predictions''
in the sense that second choices are not used during estimation, which
is based solely on first choices.} Because second choices reveal which product a consumer substitutes
to if their preferred option is unavailable, predicting them directly
validates the model\textquoteright s ability to recover substitution
patterns.

Using these experimental data, we show that both text and images
contain useful information about substitution. Intuitively, book covers
visually convey information about genre, while descriptions and reviews
capture plot details that identify similar books even within the same
genre. Our model selection algorithm chooses a review-based mixed
logit. This selected model improves counterfactual predictions of
second-choices relative to a standard attribute-based mixed logit.
In fact, the improvement relative to that model is on par with the
improvement that the attribute-based mixed logit achieves over plain
logit without random coefficients. Thus, using unstructured data alone,
our approach recovers substitution patterns substantially better than
standard attribute-based models.

We further validate our approach using observational data from 40
product categories offered on Amazon.com, including groceries, pet
food, office supplies, beauty products, electronics, video games,
and clothing. We combine online purchases from the Comscore Web Behavior
panel with product images, titles, descriptions, and reviews collected
from Amazon's product pages. We show that text and images contain
useful signals about substitution in all studied categories, highlighting
that our approach is broadly applicable across markets. We find that
text performs best in some categories where images might seem \textit{ex
ante} more relevant (e.g., clothing), and vice versa. Thus, it is
helpful to collect both text and image data even when only one seems
relevant \textit{a priori}.

Our paper contributes to the extensive literature on demand estimation.
We build on papers that model substitution through the heterogeneity
of preferences over observed product attributes \citep{BLP_1995,mcfadden2000mixed,nevo2001measuring,BLP_micro_2004}.
We show how researchers can incorporate unstructured data into these
standard models to improve estimation of substitution patterns. 
Our key contribution is to validate this approach extensively using
both experimental and observational data.

We also build on a vast computer science literature that transforms
unstructured data into lower-dimensional representations \citep{krizhevsky2012imagenet,mikolov2013efficient,goodfellow2016deep}
and a growing economics and marketing literature that applies pre-trained
deep learning models \citep{battaglia2024inference,dew2024adaptive}.
Unlike this prior work, which focuses on predicting observed outcomes,
we aim to predict substitution patterns counterfactually---a crucial
task for applications like merger simulations and optimal pricing.
Because substitution patterns are usually unobserved, we cannot apply
standard cross-validation and instead need to design an experiment
to measure them. To our knowledge, we are the first to use such an
experiment to validate demand models.\footnote{\citet{BLP_micro_2004} show that second choice data are useful for
estimating substitution patterns, and \citet{conlon2022estimating}
estimate demand from aggregate data on first-choice probabilities
and a subset of second-choice probabilities. Conceptually, our validation
approach reverses the logic of these papers: we only use first-choice
data for estimation and  reserve second choices  for model validation.
In a similar spirit to our approach, \citet{raval2022using} use hospital
closures induced by natural disasters to compare the ability of various
models to predict diversions.}

Several other papers leverage text and image data in demand estimation.
\citet{quan2019extracting} and \citet{lee2024generative} incorporate
product images and textual descriptions as mean utility shifters.
 Our approach models utility covariances and can thus capture substitution
patterns more flexibly. \citet{han2025copyright} use image data
to estimate substitution among fonts and study the effects of copyright
protection policies. We contribute by validating our approach on
ground truth substitution patterns,  comparing counterfactual prediction
performance across several embedding models and types of unstructured
data, and providing practical guidance for demand estimation. \citet{Sisodia_et_al_2022}
extract interpretable product attributes from images and use them
to estimate preferences in a choice-based conjoint experiment. Because
our approach is easier to implement and scales better across categories,
 it is preferable when researchers do not need the extracted attributes
to be interpretable.

An alternative way to learn substitution patterns is to use survey
data. For example, \citet{Dotson_et_al_2019} ask survey participants
to rate each product image and then incorporate rating correlations
as shifters of utility correlations into a demand model. Similarly,
\citet{Magnolfi_et_al_2022} elicit relative rankings from survey
participants (e.g., \textquotedblleft Product A is closer to B than
to C\textquotedblright ), generate embeddings from these rankings,
and use them in a mixed logit model. Our approach complements these
survey-based methods but has the advantage of using only widely available
text and image data, thereby avoiding the need for costly category-specific
surveys.

Our approach is well-suited for empirical applications that require
accurate and flexible estimates of product substitution. This includes
evaluating how horizontal mergers \citep{nevo2000mergers,FTC2022mergers},
new product launches \citep{hausman1994valuation,petrin2002quantifying},
corrective taxes \citep{allcott2019regressive,seiler2021impact},
and trade restrictions \citep{goldberg1995product,berry1999voluntary}
influence consumers' choices and welfare through changes in assortments
and prices. It can also be used to study how multi-product firms should
optimize prices and promotions \citep{hoch1995determinants,vilcassim1995investigating}.
The approach avoids the need to collect category-specific attributes,
making it easier to estimate demand across many categories \citep{dopper2024rising}.\footnote{\citet{dopper2024rising} remark that estimating demand across multiple
categories ``would be difficult to implement at scale because it
would require category by-category assessments about which characteristics
are appropriate to include and whether or not relevant data are available.''
Our approach can address this problem if researchers have access to
unstructured data.} Finally, this approach  might be particularly useful for estimating
demand for creative goods---such as books, art, and movies---where
unstructured data is likely to capture product differentiation more
fully than observed attributes alone.

\section{Proposed Approach \protect\label{sec:proposed_approach}}

The proposed approach involves three steps: (1) extracting embeddings
from images and texts, (2) reducing the dimensionality of these embeddings
using PCA, and (3) including the resulting principal components into
a standard attribute-based logit model with random coefficients.

\paragraph*{Step 1. Extracting Embeddings from Texts and Images}

To extract information from product images, we compute low-dimensional
representations---\textsl{embeddings}---using pre-trained deep learning
models. We use four convolutional neural networks: VGG19 \citep{Simonyan_Zisserman_2015},
ResNet50 \citep{He_et_al_2016}, InceptionV3 \citep{Szegedy_2016},
and Xception \citep{Chollet_2017}.\footnote{These models were originally trained to classify images into labeled
categories (e.g., ``cup,'' ``book,'' or ``sofa''). However,
since our goal is not to label products but to measure visual features
that distinguish them from each other, we remove the classification
layer from these models and work directly with embeddings.} The key advantage of using pre-trained models is that it reduces
the computational burden of estimation while leveraging models that
were trained on large-scale datasets and are known for their strong
performance in image classification tasks \citep{Russakovsky_et_al_2015}.
Because these models perform well at distinguishing visually similar
objects, we expect embeddings to capture the key visual features that
differentiate products. Rather than committing to a single model,
we perform model selection to determine which one best captures substitution
patterns, as described below.

We also extract information from texts, including product titles,
descriptions, and reviews. We use a bag-of-words model, which represents
text as fixed-length vectors based on word counts, as well as a variation
with a TF-IDF vectorizer.\footnote{TF-IDF assigns more weight to words that are frequent in a specific
text but rare across others, thus emphasizing unique words. This makes
it more likely that text embeddings capture distinctive words that
differentiate products from one another.} While these count models are relatively simple, we view them as useful
benchmarks because they can still detect attributes mentioned in textual
product descriptions. In addition, we employ two pre-trained deep
learning models: Universal Sentence Encoder (USE) \citep{cer_et_al_2018},
and BERT Sentence Transformer (ST) \citep{reimers-2019-sentence-bert}.\footnote{The model trained by \citet{reimers-2019-sentence-bert} is a more
efficient version of the widely used BERT network \citep{devlin2018bert}.} Both USE and ST produce semantically meaningful sentence embeddings
and achieve excellent performance on semantic textual similarity benchmarks
\citep{cer2017semeval}. As a result, they can identify when sellers
or consumers describe the same product attributes and functional benefits
using similar language, even if the exact words differ.

\paragraph*{Step 2. Generating Principal Components \protect\label{subsec:generating_pc}}

Although embeddings compress texts and images into lower-dimensional
representations, they remain high-dimensional compared to the number
of attributes typically included in demand models. For example, incorporating
512-dimensional VGG19 embeddings into a random coefficients logit
model would require prohibitively costly numerical integration over
all dimensions. We apply PCA to further reduce dimensionality \citep{Backus_et_al_2021}.

PCA is appealing for several reasons. Because embeddings are trained
on general-purpose datasets to classify images into broad categories,
they contain variation that distinguishes each category from one another
(e.g., laptops from tablets). By applying PCA within each category,
we downweight this common variation and focus on differences across
products within that category. Additionally, the resulting principal
components are orthogonal to each other, which avoids multicollinearity
issues and thus simplifies estimation of logit demand models with
random coefficients.

\paragraph*{Step 3. Including Principal Components into a Demand Model \protect\label{subsec:model_selection}}

We include principal components into a standard mixed logit model
\citep{BLP_1995,BLP_micro_2004}, estimating a separate random coefficient
for each included component. Since each deep learning model and data
type produces different principal components, a key question is which
to include in the demand model.\footnote{While we could select the specification that best matches observed
second choices in our experiment, this strategy is unavailable to
most researchers who usually only have first-choice data. We thus
propose a model selection algorithm that relies solely on first choices.} We perform model selection using in-sample $AIC$, which balances
model fit and complexity and can be interpreted as approximating expected
out-of-sample predictive performance \citep{akaike1998information}.
In our experimental results, in-sample $AIC$ strongly correlates
with counterfactual performance, justifying its use for model selection
(see Section \ref{subsec:second_choice_validation} for a discussion
and comparison to $BIC$ and cross-validation).

Algorithm 1 describes our model selection procedure. We define the
\textit{candidate set} as the variables for which random coefficients
may be included. In our applications,  this set includes price and
the first $P$ principal components.  We first estimate the plain
logit model without random coefficients ($K=0$) and record its $AIC$.
For each $K\geq1$, we estimate all models with random coefficients
on $K$ candidate variables and retain the model with the lowest $AIC$.
If this \textit{$AIC$} is lower than that of the previously selected
model, we update the selected specification and increase $K$; otherwise,
we stop and select the previous model. We increase $K$ until $AIC$
no longer improves. We repeat this procedure for each specification---defined
by data type and embedding model (e.g., \textit{USE Reviews})---and
choose the specification with the lowest overall $AIC$.

This algorithm avoids costly combinatorial search over all possible
specifications. At the same time, it ensures that we do not rely too
heavily on the default ordering of principal components, which reflects
the explained variance of embeddings but need not match how predictive
they are of preferences. For example, substitution patterns may be
driven by the second and fourth principal components. In this case,
our approach may select a model with random coefficients on those
components, but not on the first or third.

\begin{table}[t]
\centering{}%
\begin{tabular}{V{\linewidth}V{\linewidth}}
\hline 
\multicolumn{2}{l}{{\small\textbf{Algorithm 1:}}{\small{} Model Selection}}\tabularnewline
\hline 
\multicolumn{2}{V{\linewidth}}{{\small\textbf{Input:}}{\small\par}

- {\small Products $j=1,\dots,J$}{\small\par}

{\small - Specifications $m=1,\dots,M$ (all combinations of unstructured
data and pre-trained text or image models)}{\small\par}

{\small - Maximum number of principal components $P$ (researcher's
choice)}{\small\par}

{\small - Demand model with utility: $u_{ij}=\delta_{j}+\alpha_{i}\text{price}_{ij}+\theta_{i}'PC_{j}+\varepsilon_{ij}$
($\delta_{j}$ = product fixed effects)}}\tabularnewline
\hline 
{\small 1:\vspace{4pt}
} & {\small\textbf{For }}{\small each specification $m$ }{\small\textbf{do}}\tabularnewline
{\small 2:\vspace{4pt}
} & {\small$\quad$Extract embeddings $e^{(m)}=(e_{1}^{(m)},\dots,e_{J}^{(m)})$}\tabularnewline
{\small 3:\vspace{4mm}
} & {\small$\quad$Extract principal components $PC_{1}^{(m)},\dots,PC_{P}^{(m)}$
from embeddings $e^{(m)}$ using PCA}\tabularnewline
{\small 4:\vspace{4pt}
} & {\small$\quad$Initialize model selection at $K\leftarrow0$ (zero
random coefficients)}\tabularnewline
{\small 5:} & {\small$\quad$Set $\text{BestSpec}^{(m)}$ as }{\small\textit{Plain
Logit}}{\small{} and $\text{BestAIC}^{(m)}$ as the corresponding $AIC$}\tabularnewline
{\small 6:\vspace{4pt}
} & {\small$\quad$}{\small\textbf{while}}{\small{} $\text{BestAIC}^{(m)}$
decreases }{\small\textbf{do}}\tabularnewline
{\small 7:\vspace{4pt}
} & {\small$\quad$$\quad$$K\leftarrow K+1$}\tabularnewline
{\small 8:\vspace{4pt}
} & {\small$\quad$$\quad$Estimate mixed logit models with all possible
combinations of random coefficients}\tabularnewline
 & {\small$\quad$$\quad$ on subsets of $K$ variables from the candidate
set $\mathcal{R}=\{\text{price, P\ensuremath{C_{1}^{(m)}}, \ensuremath{\dots}, P\ensuremath{C_{P}^{(m)}}}\}$}\tabularnewline
{\small 9:\vspace{4pt}
} & {\small$\quad$$\quad$Find subset $R_{K}^{*}\subseteq\mathcal{R}$
that minimizes $AIC$ at $AIC_{K}^{*}$}\tabularnewline
{\small 10:\vspace{4pt}
} & {\small$\quad$$\quad$}{\small\textbf{if }}{\small$AIC_{K}^{*}<\text{BestAIC}^{(m)}$
}{\small\textbf{do}}\tabularnewline
{\small 11:\vspace{4pt}
} & {\small$\quad$$\quad$$\quad$$\text{BestSpec}^{(m)}\leftarrow R_{K}^{*}$
: update the best specification}\tabularnewline
{\small 12:\vspace{4pt}
} & {\small$\quad$$\quad$$\quad$$\text{BestAIC}^{(m)}\leftarrow AIC_{K}^{*}$
: update the lowest $AIC$}\tabularnewline
{\small 13:\vspace{4pt}
} & {\small$\quad$$\quad$}{\small\textbf{end if}}\tabularnewline
{\small 14:\vspace{4pt}
} & {\small$\quad$}{\small\textbf{end while}}\tabularnewline
{\small 15:\vspace{4pt}
} & {\small\textbf{end for}}\tabularnewline
{\small 16:\vspace{4pt}
} & {\small Choose the best-fitting specification $m^{*}=\arg\min_{m}\text{BestAIC}^{(m)}$}\tabularnewline
\hline 
\end{tabular}
\end{table}

\paragraph*{Why Does This Approach Work?}

The key feature of our approach is that it uses text and images as
proxies for product attributes that shape substitution. Take tablets,
for example: product titles may reveal brand, screen size, and camera
resolution (e.g., ``Apple iPad 10.2-inch 12MP camera''); seller
descriptions may highlight whether the tablet is suitable for drawing
or gaming; and consumer reviews may mention that a tablet is durable
and child-friendly. Similarly, photos may showcase design features
like color and casing style. Our approach extracts this information
in an automated way and uses it to estimate substitution patterns.

We interpret the relationship between embeddings and substitution
as correlational. In particular, our approach does not assume that
consumers directly consider product images or textual descriptions
when making choices.   Under this interpretation, our  approach
is well-suited for counterfactuals where embeddings can be held fixed,
such as optimal pricing or merger simulations.  By contrast, it is
less appropriate for settings in which counterfactual changes that
may alter product design or positioning and, in turn, the embeddings
themselves. In such cases, researchers are required to estimate the
causal effect of  embeddings on demand. Section \ref{subsec:counterfactuals}
provides  additional discussion.

\paragraph*{Advantages Over Standard Methods \protect\label{subsec:advantages_of_approach}}

Our approach has several advantages relative to the standard attribute-based
methods. First, researchers typically select a limited set of attributes
based on their prior knowledge of the market, often those that are
easiest to quantify, or rely on attributes supplied by data providers.
Our approach avoids these subjective choices by automatically extracting
information about product substitution from unstructured data. Second,
this approach captures visual design and functional benefits, which
may drive substitution but are difficult to capture through observed
attributes. Lastly, we circumvent the need to collect category-specific
attributes, which makes our approach more scalable. This is valuable
for researchers seeking to estimate demand and study competition and
pricing across many product categories. 

When some data on observed attributes are available, researchers can
incorporate them into our approach. For example, they can expand the
set of candidate variables in Algorithm 1 to include the observed
attributes themselves or, when the number of attributes is large,
their principal components.  In our experiment, adding observed
attributes does not improve counterfactual predictions, indicating
that the information in the attributes is subsumed by the unstructured
data. See Sections \ref{subsec:second_choice_validation} and \ref{subsec:practitioners_guide}
for details.

\section{Validation with Experimental Data \protect\label{sec:choice_experiment}}

Next, we validate our approach empirically. A key challenge is that
substitution patterns are unobserved, so there is typically no ground
truth against which to test model predictions. To address this challenge,
we design a choice experiment that provides such a ground-truth measure
of substitution.

\subsection{Experiment Design}

\paragraph*{Choice Tasks.}

Each participant completes two choice tasks. In the first choice task,
a participant chooses one book from a set of ten alternatives. We
instruct participants to choose books they would purchase if faced
with this selection in a real bookstore. We limit the number of options
to ten to ensure participants pay attention to most options, which
allows us to abstract away from limited consideration. After a participant
selects a book, we remove it from the list. The participant then proceeds
to a second choice task, where they choose among the remaining nine
books.

\begin{figure}
\begin{centering}
{\small\includegraphics[width=0.8\textwidth]{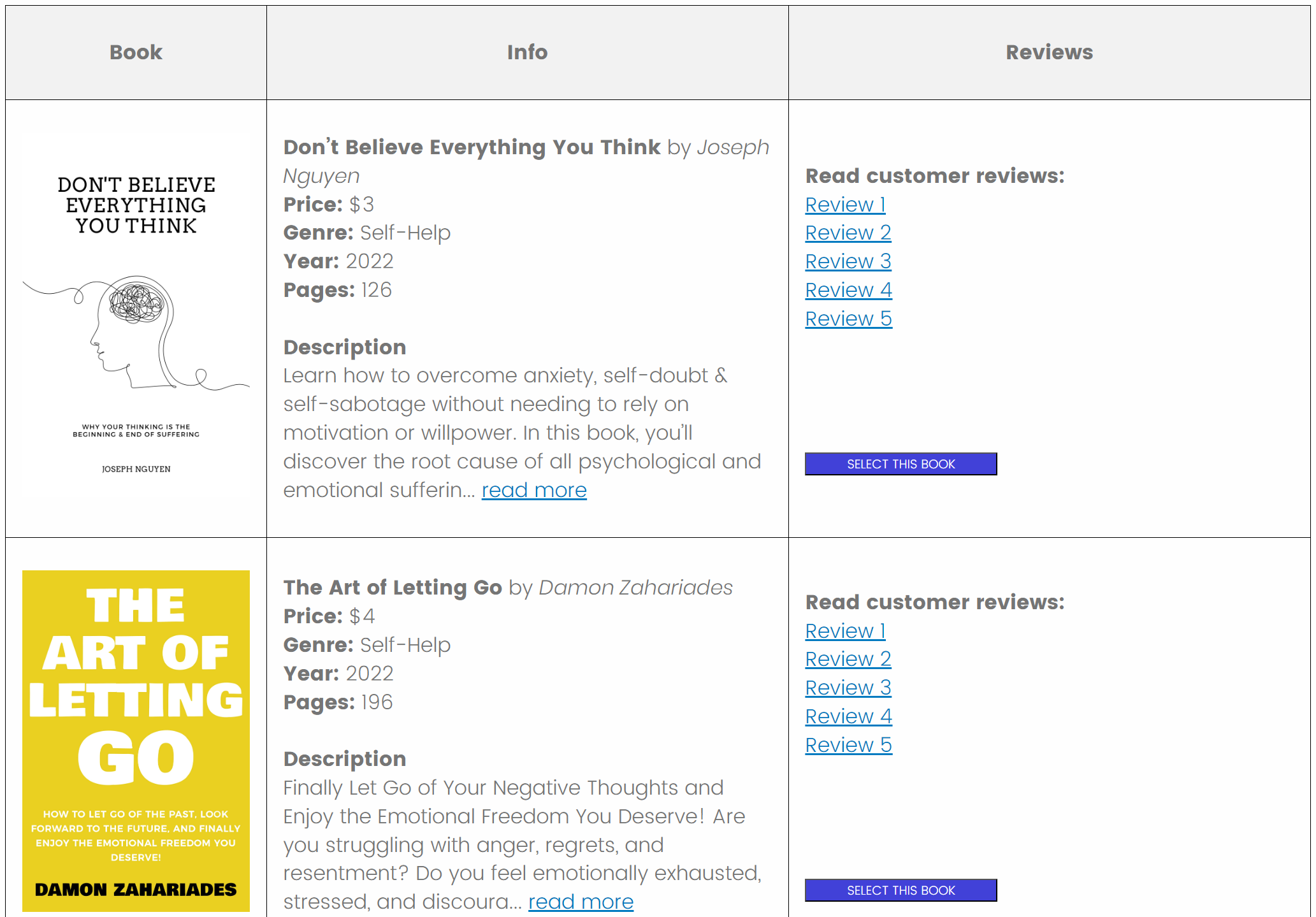}}{\small\par}
\par\end{centering}
\caption{{\footnotesize\protect\label{fig:choice_task}}{\footnotesize\textbf{Example
of a choice task in our experiment. }}{\footnotesize The screenshot
displays the top portion of the page as it appeared to participants.}}
\end{figure}

Figure \ref{fig:choice_task} shows the choice task as presented to
participants. To give all models a fair chance at capturing substitution,
we display attributes (author, year, genre, pages), cover images,
and texts (plot description and five reviews), collected from Amazon
product pages.\footnote{We collect the first five reviews displayed on each product detail
page at the time of data collection.} We take all books from Amazon's bestseller lists, sampling them in
a way that does not bias our model comparisons in favor of any particular
specification (see Appendix \ref{sec:appendix_book_selection} for
details). Lastly, we randomize book rankings and prices across participants,
keeping them fixed across choice tasks for the same participant.\footnote{Prices were drawn from a discrete uniform distribution ranging from
\$3 to \$7.} This variation allows us to cleanly estimate substitution patterns
from only first choices and reserve second-choice data for validation
\citep{berry2024nonparametric}.

The book category is well-suited for our analysis because participants
are likely to consider both structured attributes (e.g., genre) and
unstructured information (e.g., plot descriptions). It is also unclear
a priori whether substitution patterns are better predicted by images
or texts---while texts describe rich plot details, cover images contain
clear cues of genres, and whether the book is fiction or non-fiction
(see Section \ref{subsec:second_choice_validation} for further discussion).

\paragraph*{Recruitment and Sample selection.}

We recruited 10,775 participants from the online platform Prolific
between June 14 and June 27, 2024. This sample size is close to our
pre-registered target of 10,000, determined from power calculations
in a pilot experiment.\footnote{The pre-registration document can be accessed at \href{https://tinyurl.com/ekjp4pfp}{https://tinyurl.com/ekjp4pfp}.
At the time of pre-registration, we planned to estimate a different
choice model---the pairwise combinatorial logit (PCL) model from
\citet{Koppelman_Wen_2000}---which generalizes the logit model by
allowing utility correlations across product pairs. We switched to
a mixed logit model at a later stage because we found it to perform
better. We adhered to our pre-registered protocol for all key aspects
of survey design, including sample size, sample selection, and choice
task construction.} Following the pre-registered sample selection criteria, we excluded
14\% participants who failed comprehension questions, did not complete
the survey, or spent less than one minute on the study, leaving a
final sample of 9,265 participants.

Because the choice tasks were hypothetical and not incentivized, it
is important to verify that participants made meaningful selections.
In Appendix \ref{sec:sanity_checks}, we show that participants did
not rush through the survey, responded to changes in book rankings
and prices, and made choices consistent with their self-reported genre
preferences.

\subsection{Model Specifications and Estimation Results \protect\label{subsec:model_specifications}}

We compare our approach against two benchmark models: the plain logit
model and a mixed logit model with attributes. All three models fall
into a framework where the indirect utility of participant $i$ from
book $j$ is

\begin{equation}
u_{ij}=\beta_{i}'x_{j}+\theta_{i}'PC_{j}+\gamma\cdot\text{rank}_{ij}+\alpha_{i}\cdot\text{price}_{ij}+\delta_{j}+\varepsilon_{ij}.\label{eq:mixed_logit_utility}
\end{equation}
Here $x_{j}$ are observed book attributes: genre dummies, publication
year, and length in pages---all attributes participants observe in
the choice tasks; $PC_{j}$ are principal components extracted from
embeddings, $\text{price}_{ij}$ and $\text{rank}_{ij}$ are the price
and position of book $j$ in participant $i$'s choice task, $\delta_{j}$
are product fixed effects, and $\varepsilon_{ij}$ are i.i.d. taste
shocks following a Type I Extreme Value distribution. We include $\text{rank}_{ij}$
in the model because our experiment induces random variation in the
placement of books.\footnote{We do not include a random coefficient on $\text{rank}_{ij}$ in any
model specification because our focus is on estimating substitution
patterns using observed characteristics or principal components.} This variation gives us an additional exogenous shifter that helps
us identify substitution patterns, analogous to price variation.

We estimate three demand models:
\begin{enumerate}
\item \textbf{Mixed Logit with Principal Components.} We include principal
components $PC_{j}$, omit observed attributes $x_{j}$, and assume
$\alpha_{i}\sim N(\bar{\alpha},\sigma_{\alpha})$ and $\theta_{i}\sim N(0,\Sigma_{\theta})$
with diagonal $\Sigma_{\theta}$.\footnote{Since the principal components do not vary over time, the mean of
$\theta_{i}$ is absorbed by the product fixed effects $\delta_{j}$.
The same holds for the mean of $\beta_{i}$ in the mixed logit model
with attributes. In addition, we include the mean price coefficient
$\bar{\alpha}$ in all specifications, regardless of whether the random
coefficient on price is selected by the model selection algorithm.} We perform model selection as detailed in Algorithm 1 in Section
\ref{subsec:model_selection}, letting the set of candidate variables
include price and the first $P=6$ principal components.\footnote{The first six principal components together explain 70-80\% of the
variance in embeddings (Appendix Figure \ref{fig:pc_explained_variance}).}
\item \textbf{Mixed Logit with Attributes.} We include observed attributes
$x_{j}$, omit principal components $PC_{j}$, and assume $\alpha_{i}\sim N(\bar{\alpha},\sigma_{\alpha})$
and $\beta_{i}\sim N(0,\Sigma_{\beta})$ with diagonal $\Sigma_{\beta}$.
We choose the lowest $AIC$ specification among all possible combinations
of random coefficients on price, publication year, length in pages,
and genre.
\item \textbf{Plain Logit.} We include neither attributes $x_{j}$ nor principal
components $PC_{j}$, estimating only a constant price coefficient
$\alpha_{i}=\alpha$, rank coefficient $\gamma$, and fixed effects
$\delta_{j}$.
\end{enumerate}
Table \ref{tab:aic_fit} compares in-sample $AIC$. The best mixed
logit specification with attributes includes random coefficients on
the number of pages and year, reducing $AIC$ by $16.0$ relative
to plain logit. The best model with unstructured data, \textit{Reviews
USE}, puts random coefficients on the first two principal components.
This model fits better than any other considered specification, reducing
$AIC$ by $24.8$ relative to plain logit and by $8.8$ relative to
the best attribute-based logit. While this superior fit is reassuring,
it does not guarantee better counterfactual performance. We therefore
turn to predicting second choices counterfactually.

\begin{table}
\begin{centering}
	\begin{centering}
\begin{tabular}{>{\raggedright}p{7.5cm}>{\raggedright}p{3.5cm}>{\centering}p{1.5cm}>{\centering}p{1.5cm}}
\hline
{\footnotesize\textbf{Model}} & {\footnotesize\textbf{Random Coefficients}} & {\footnotesize\textbf{$AIC$}} & {\footnotesize\textbf{$\Delta AIC$}}\tabularnewline
\hline
{\footnotesize Plain Logit} & {\footnotesize None} & {\footnotesize 41006.7} & {\footnotesize 0.0}\tabularnewline
{\footnotesize Mixed Logit with Attributes} & {\footnotesize Pages and Year} & {\footnotesize 40990.7} & {\footnotesize -16.0}\tabularnewline
{\footnotesize Mixed Logit with Images (InceptionV3)} & {\footnotesize Price, PC1, and PC6} & {\footnotesize 40990.5} & {\footnotesize -16.2}\tabularnewline
{\footnotesize Mixed Logit with Titles (ST)} & {\footnotesize PC1 and PC5} & {\footnotesize 40992.3} & {\footnotesize -14.4}\tabularnewline
{\footnotesize Mixed Logit with Descriptions (USE)} & {\footnotesize PC1 and PC5} & {\footnotesize 40986.4} & {\footnotesize -20.3}\tabularnewline
{\footnotesize Mixed Logit with Reviews (USE)} & {\footnotesize PC1 and PC2} & {\footnotesize 40981.9} & {\footnotesize -24.8}\tabularnewline
\hline
\end{tabular}
\par\end{centering}
\par\end{centering}
\caption{{\small\textbf{\protect\label{tab:aic_fit}Comparison of models in
terms of in-sample $AIC$ on first choices. }}{\small The second column
shows variables that have random coefficients in the selected specification.
The last column shows the }{\small\textbf{$AIC$}}{\small{} reduction
relative to plain logit.}}
\end{table}

\subsection{Validation Using Second-Choice Data \protect\label{subsec:second_choice_validation}}

 We assess how well different demand models predict second choices
counterfactually. Because second choices reveal which books participants
choose when their preferred option is unavailable, predicting them
accurately requires a model that captures substitution patterns well.
Further, predicting second choices is directly relevant in antitrust
where second-choice \textit{diversion ratios} are often used as measures
of substitutability and price competition \citep{conlon2021empirical}.

We first estimate each model via Maximum Likelihood Estimation (MLE)
using only first choices. We then use the estimated model to predict
counterfactual second-choice diversion ratios, $\ensuremath{\hat{s}_{j\rightarrow k}}$,
defined as the probability that the participant chooses product $k$
in the second choice task conditional on having chosen book $j$ in
the first choice task. We compare these predictions with diversions
observed in the data, $\ensuremath{s_{j\rightarrow k}}$, computing
$RMSE$ as:

\begin{equation}
RMSE=\sqrt{\frac{1}{J(J-1)}\sum_{j}\sum_{k\neq j}\left(s_{j\rightarrow k}-\hat{s}_{j\rightarrow k}\right)^{2}}\label{eq:rmse_equation}
\end{equation}

\noindent Since we do not use second-choice data in estimation, lower
$RMSE$ indicates better performance at counterfactual predictions.
Appendix \ref{sec:rmse_computations} provides further details on
how we compute diversions $\ensuremath{s_{j\rightarrow k}}$ and $\ensuremath{\hat{s}_{j\rightarrow k}}$.

\begin{figure}[!t]
\centering{}{\small\includegraphics[width=0.8\textwidth]{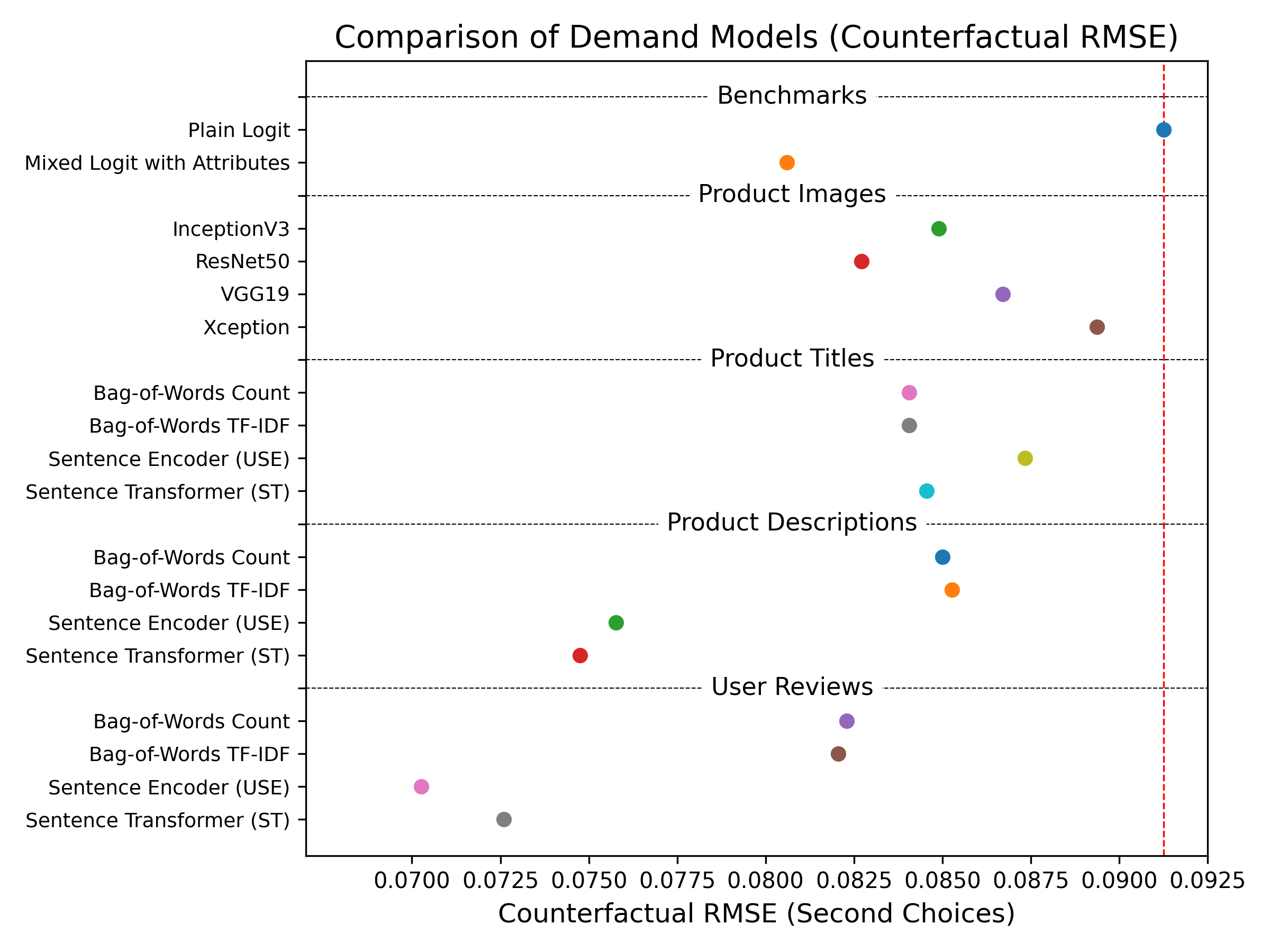}}\caption{{\small\textbf{\protect\label{fig:model_comparison_rmse} Comparison
of models in terms of counterfactual $RMSE$ on second choices.}}{\small{}
The two benchmarks in the first panel are the plain logit without
random coefficients and the lowest-$AIC$ mixed logit with random
coefficients on observed attributes. The remaining specifications
correspond to mixed logit models with random coefficients on principal
components extracted from image or text embeddings.}}
\end{figure}

Figure \ref{fig:model_comparison_rmse} summarizes the validation
results, while Appendix Table \ref{tab:validation_numbers} reports
$AIC$ and $RMSE$ values for all estimated specifications. The top
panel in Figure \ref{fig:model_comparison_rmse} shows two benchmarks:
the plain logit and the lowest-$AIC$ mixed logit with attributes.
The mixed logit with attributes reduces RMSE by 11.7\%, while our
best-fitting \textit{Review USE} model reduces RMSE by 23\% relative
to plain logit, significantly outperforming the mixed logit with attributes.
This result illustrates the value of our approach. We chose books
for this study expecting observed attributes like genre to predict
substitution well. Yet, by using unstructured data, we can match the
counterfactual performance of the mixed logit with attributes and
even further reduce $RMSE$ by 14\%. This shows that our approach
recovers substitution patterns better than standard attribute-based
methods in this dataset.

Although the model with product reviews performs best, in general,
performance varies widely across specifications. Below, we discuss
why some specifications recover substitution patterns better than
others.

\paragraph*{Images}

\begin{figure}
\begin{centering}
\begin{tabular}{c}
\begin{cellvarwidth}[t]
\centering
{\large\textbf{Mystery Books}}{\large\par}

\begin{tabular}{cccc}
\begin{cellvarwidth}[t]
\centering
{\small\includegraphics[width=3cm,totalheight=4cm]{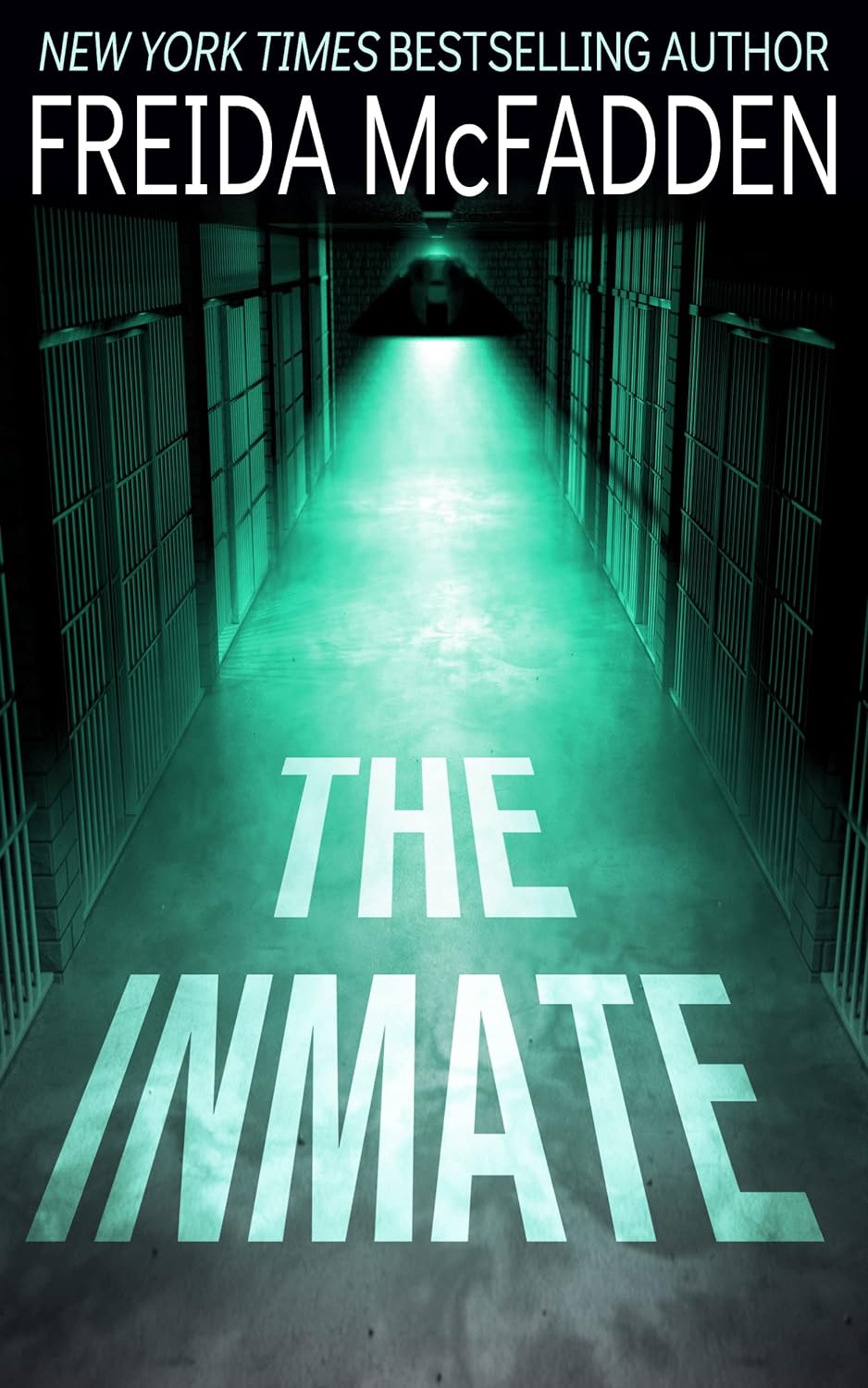}}{\small\par}

{\small ``The Inmate''}
\end{cellvarwidth} & \begin{cellvarwidth}[t]
\centering
{\small\includegraphics[width=3cm,totalheight=4cm]{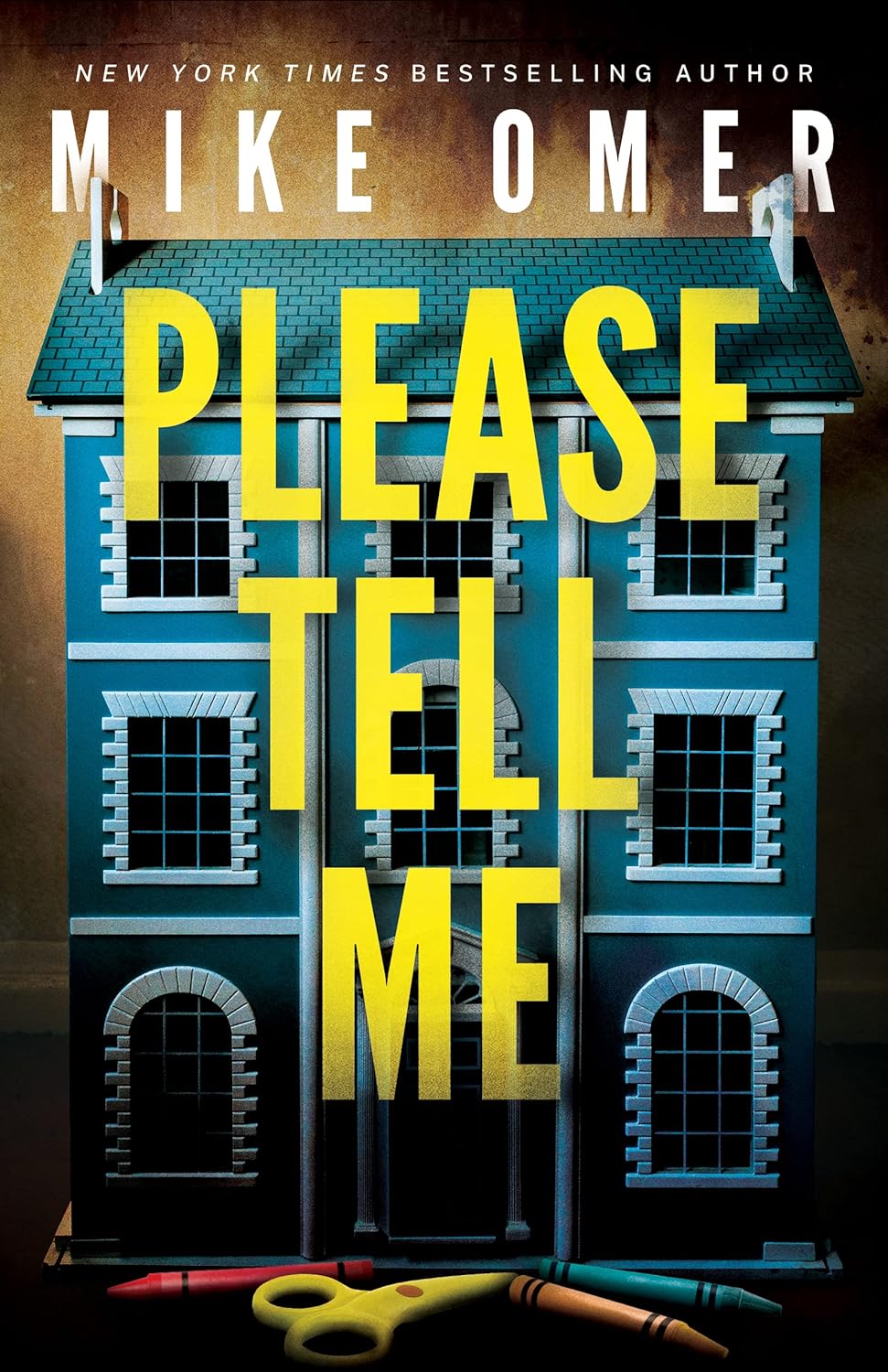}}{\small\par}

{\small ``Please Tell Me''}
\end{cellvarwidth} & \begin{cellvarwidth}[t]
\centering
{\small\includegraphics[width=3cm,totalheight=4cm]{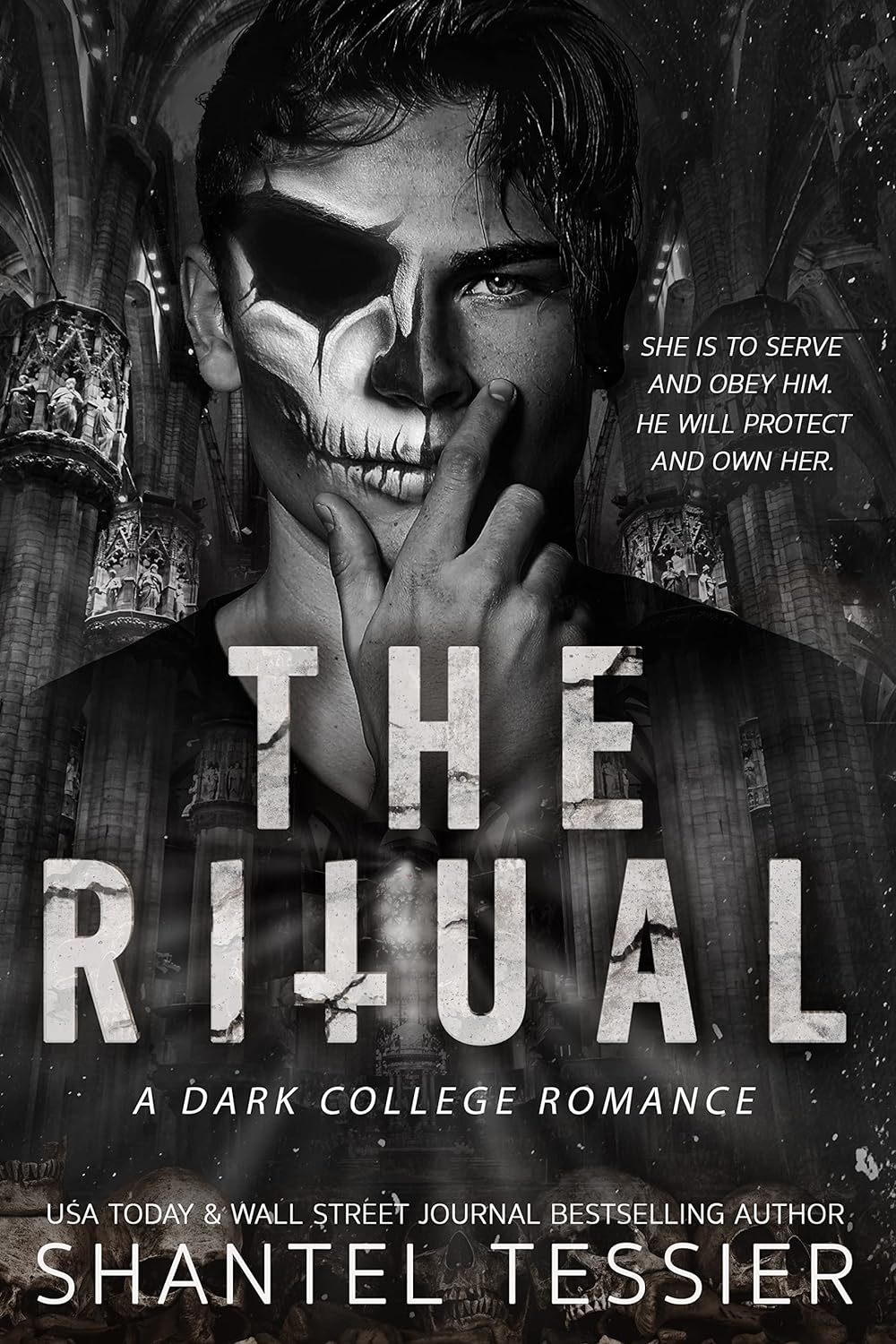}}{\small\par}

{\small ``The Ritual''}
\end{cellvarwidth} & \begin{cellvarwidth}[t]
\centering
{\small\includegraphics[width=3cm,totalheight=4cm]{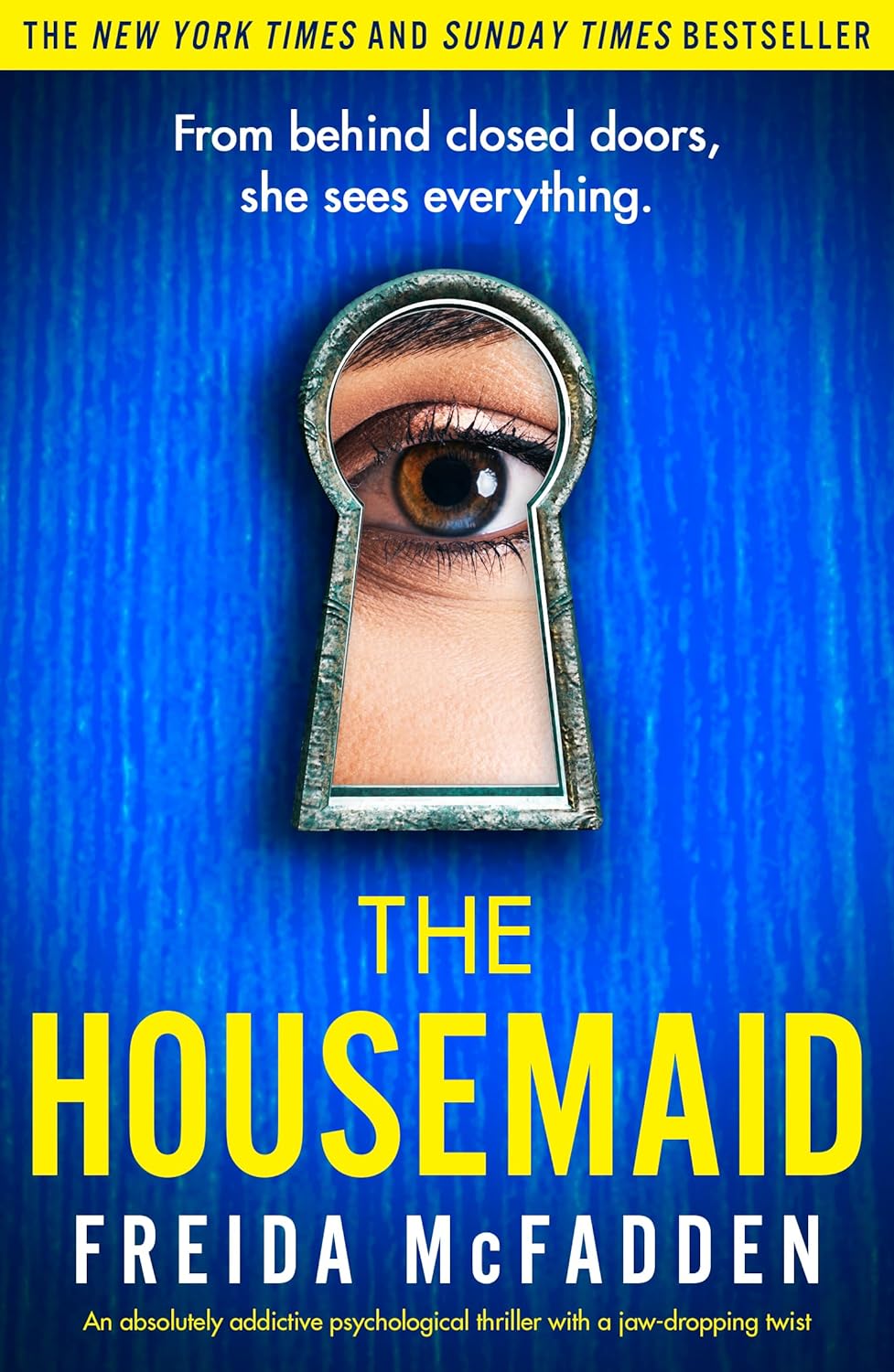}}{\small\par}

{\small ``The Housemaid''}
\end{cellvarwidth}\tabularnewline
\end{tabular}

\vspace{10pt}
\end{cellvarwidth}\tabularnewline
\begin{cellvarwidth}[t]
\centering
{\large\textbf{Fantasy Books}}{\large\par}

\begin{tabular}{ccc}
\begin{cellvarwidth}[t]
\centering
{\small\includegraphics[width=3cm,totalheight=4cm]{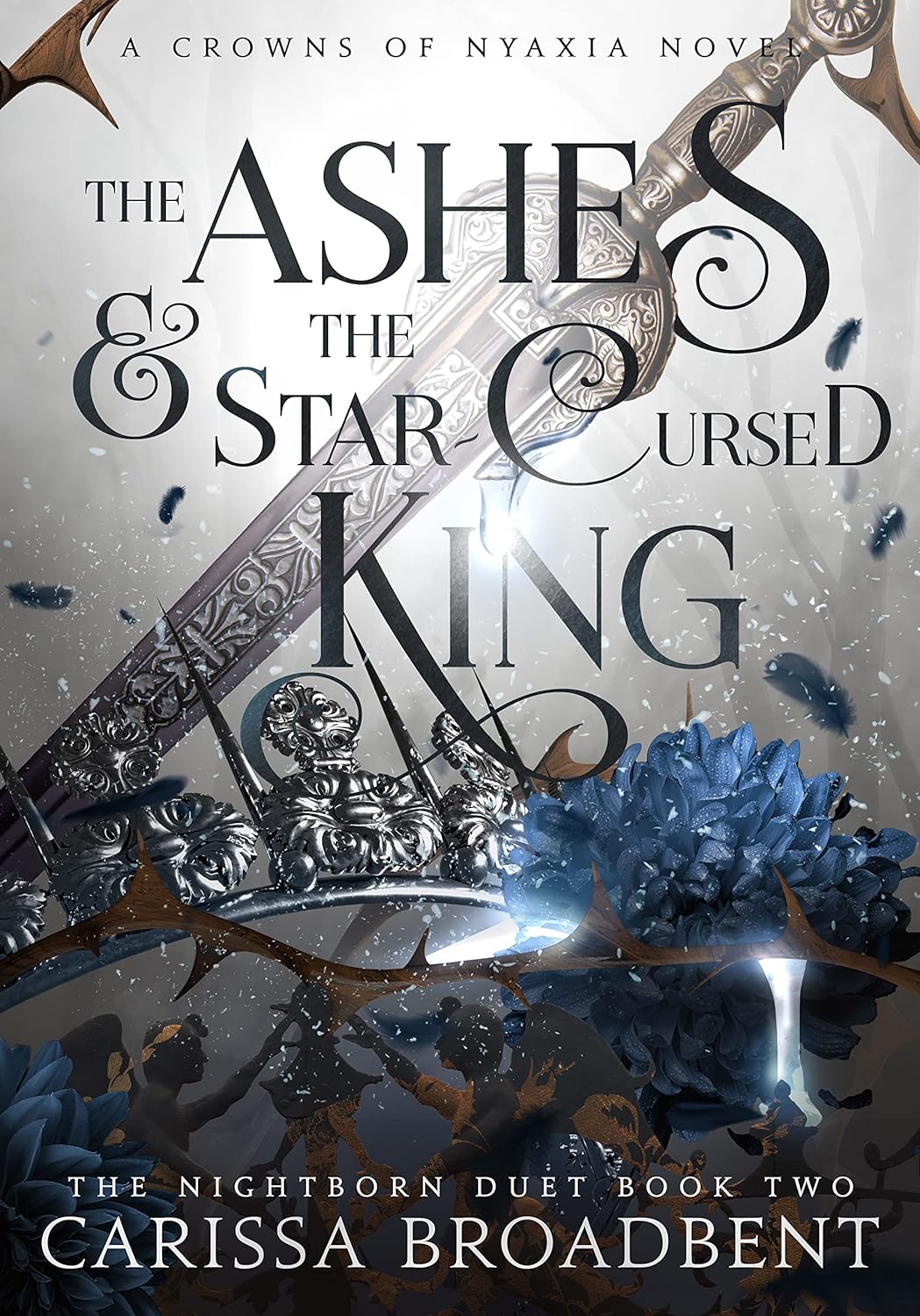}}{\small\par}

{\small ``The Ashes \& The Star}{\small\par}

{\small Cursed King''}
\end{cellvarwidth} & \begin{cellvarwidth}[t]
\centering
{\small\includegraphics[width=3cm,totalheight=4cm]{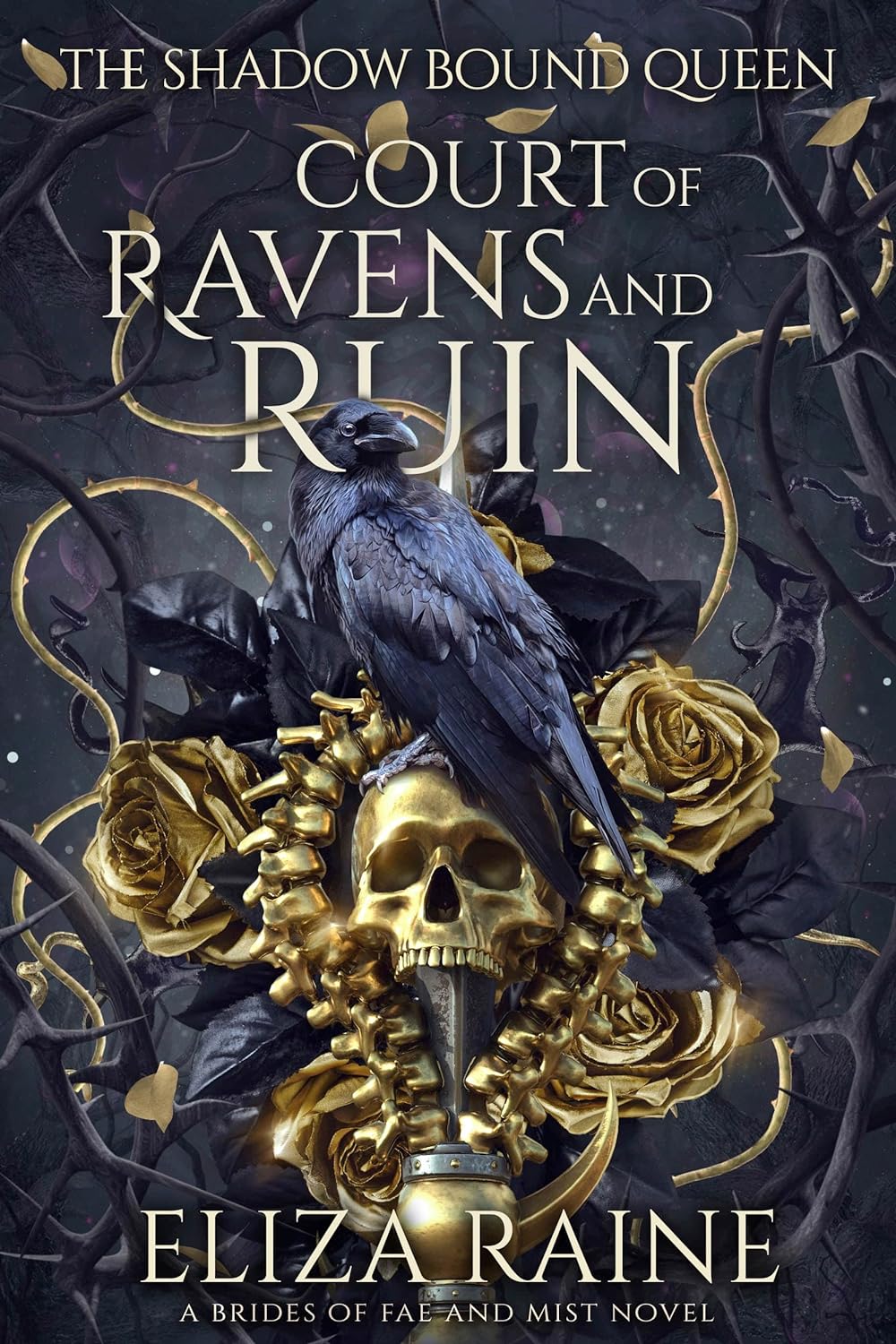}}{\small\par}

{\small ``Court of Ravens}{\small\par}

{\small and Ruin''}
\end{cellvarwidth} & \begin{cellvarwidth}[t]
\centering
{\small\includegraphics[width=3cm,totalheight=4cm]{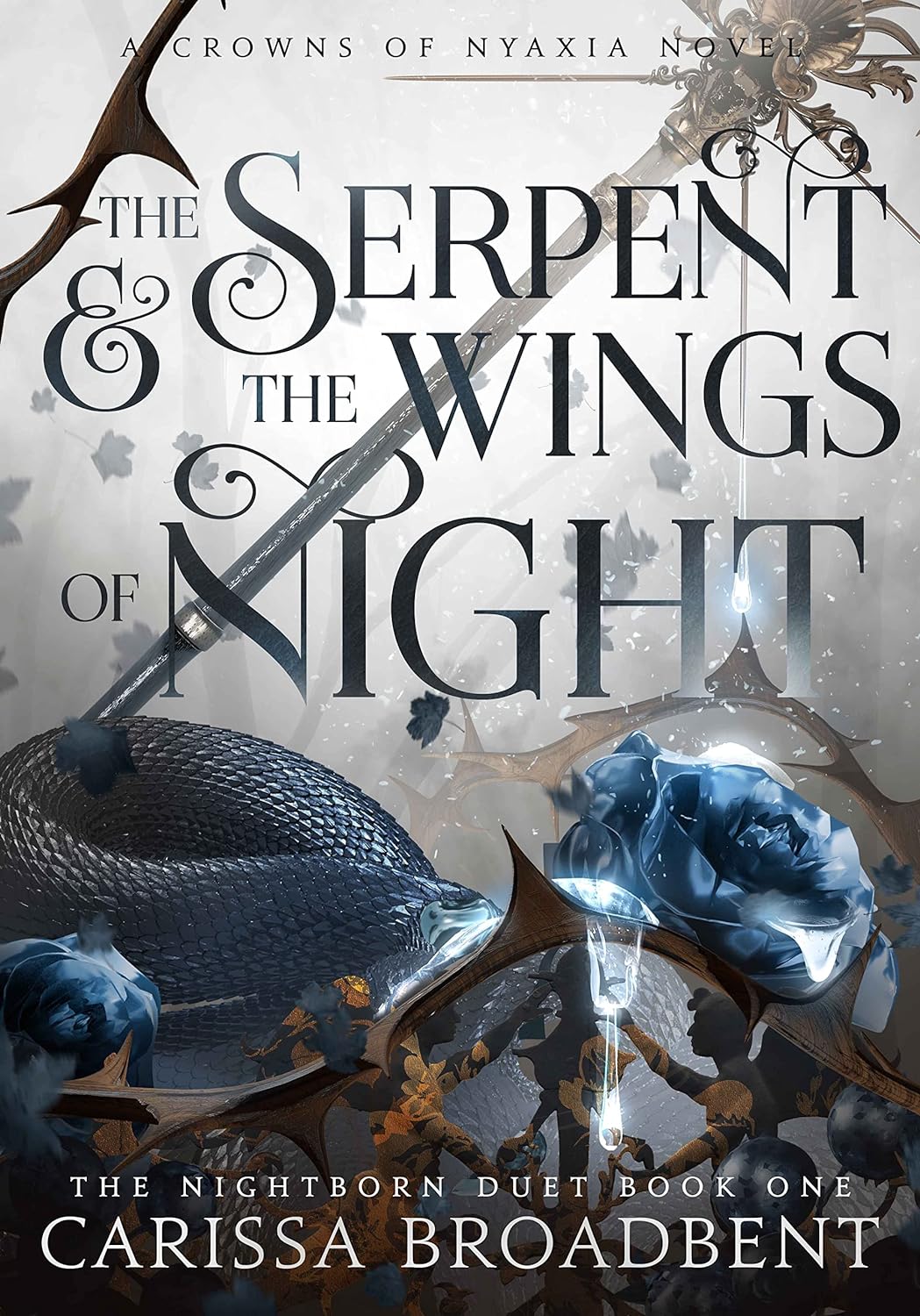}}{\small\par}

{\small ``The Serpent \&}{\small\par}

{\small The Wings of Night''}
\end{cellvarwidth}\tabularnewline
\end{tabular}

\vspace{10pt}
\end{cellvarwidth}\tabularnewline
\begin{cellvarwidth}[t]
\centering
{\large\textbf{Self-Help Books}}{\large\par}

\begin{tabular}{ccc}
\begin{cellvarwidth}[t]
\centering
{\small\includegraphics[width=3cm,totalheight=4cm]{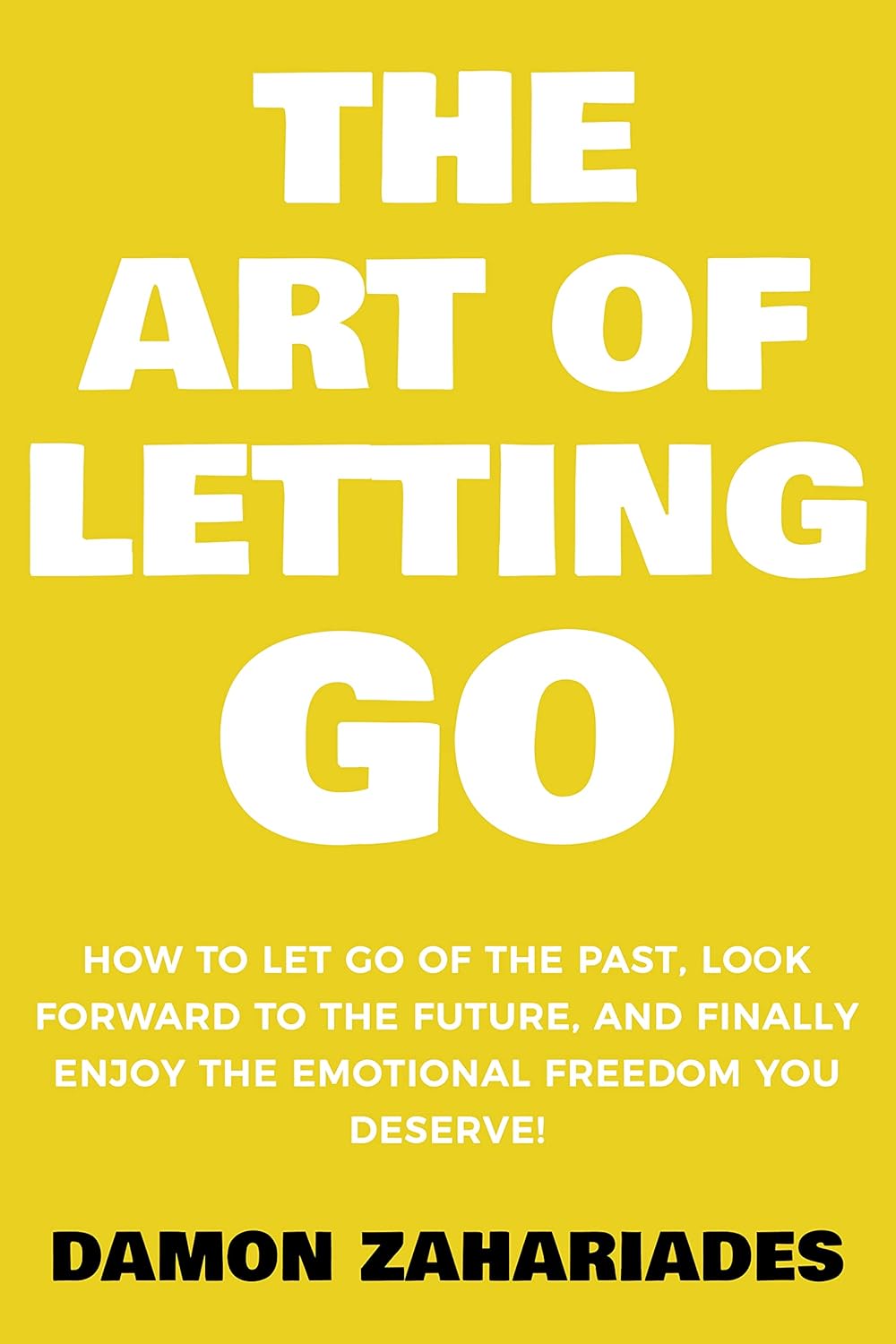}}{\small\par}

{\small ``The Art of}{\small\par}

{\small Letting Go''}
\end{cellvarwidth} & \begin{cellvarwidth}[t]
\centering
{\small\includegraphics[width=3cm,totalheight=4cm]{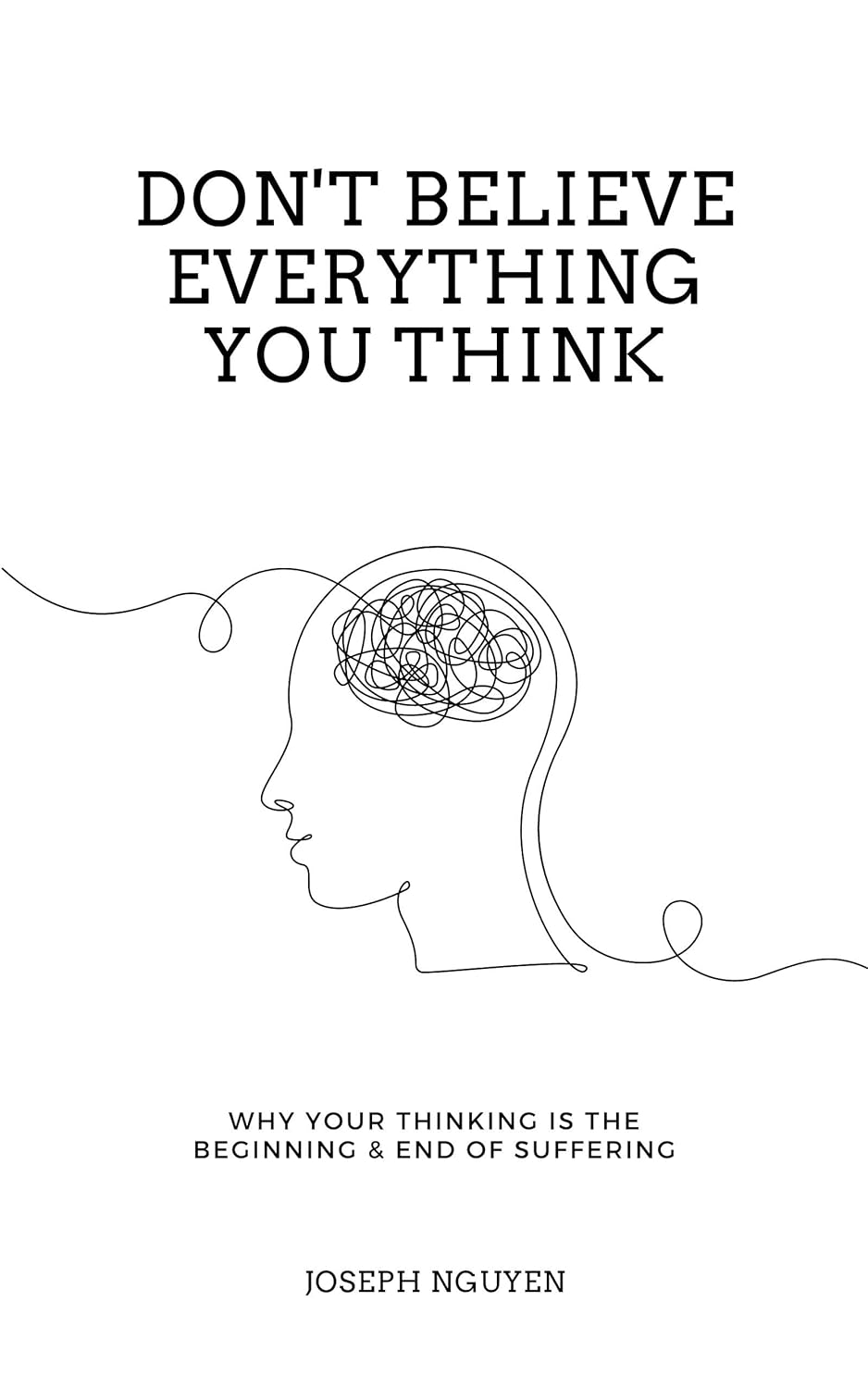}}{\small\par}

{\small ``Don't Believe}{\small\par}

{\small Everything}{\small\par}

{\small You Think''}
\end{cellvarwidth} & \begin{cellvarwidth}[t]
\centering
{\small\includegraphics[width=3cm,totalheight=4cm]{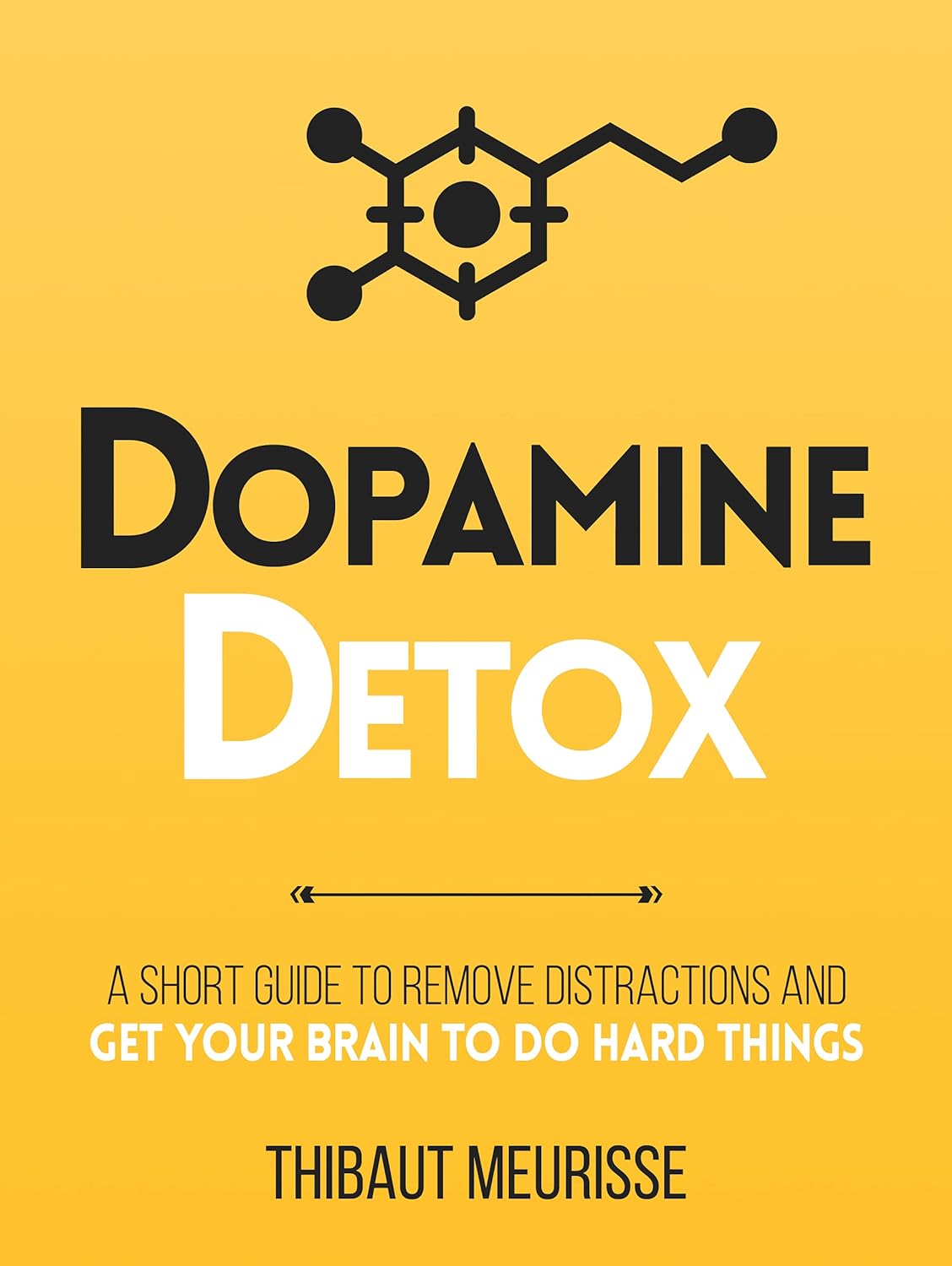}}{\small\par}

{\small ``Dopamine}{\small\par}

{\small Detox''}
\end{cellvarwidth}\tabularnewline
\end{tabular}

\vspace{10pt}
\end{cellvarwidth}\tabularnewline
\end{tabular}
\par\end{centering}
\caption{{\small\textbf{\protect\label{fig:book_covers} Ten books used in
our experiment.}}}
\end{figure}

All four image models outperform the plain logit model, with the selected
(i.e., lowest-$AIC$) model, InceptionV3, reducing $RMSE$ by 7.0\%
relative to plain logit. To understand why images predict substitution,
consider the book covers used in our study (see Figure \ref{fig:book_covers}).
Within the same genre, book covers often share similar design elements.
For example, the covers of all three fantasy books use dark, muted
color palettes with metallic accents, include symbolic elements such
as skulls and swords that convey danger or peril, and feature natural
objects like twisted vines and golden roses. Similarly, the self-help
books have minimalistic layouts and consistent color schemes, such
as black-and-white text on yellow backgrounds. Thus, image embeddings
partly encode books' genres, which correlate with substitution patterns.

\paragraph*{Texts}

All text models outperform the plain logit, and their relative performance
highlights several notable patterns.

First, performance improves as we move from simple bag-of-words models
to the more advanced USE and ST models. This is particularly true
for descriptions and reviews, which contain detailed information about
book plots but do not always use the same words or phrases. Therefore,
extracting substitution patterns from text requires natural language
models that can accurately measure semantic similarity.

Second, performance improves as we include richer text data. For
instance, relative to the plain logit, the USE model reduces $RMSE$
by 4.3\% using titles, 17\% using descriptions, and 23\% using reviews.
This makes intuitive sense. Take mystery books, for example. Titles
may contain subtle genre cues, hinting at characters in confined situations
(\textsl{Housemaid}, \textit{Inmate}). Descriptions reveal further
plot details, such as hidden identities, past relationships, and betrayals.
Lastly, customer reviews contain the richest information about a book's
style and pacing: for instance, readers may praise cliffhangers and
intriguing twists, or critique pacing issues like slow starts or rushed
endings.

\paragraph*{Extension: Combining Data Types}

If text, images, and attributes provide distinct signals of substitution,
combining them might improve fit and counterfactual performance. To
explore this, we start from the selected \textit{Review USE} model
and attempt to extend it in two ways: (1) by adding a combination
of observed product attributes with random coefficients, or (2) by
adding a combination of principal components from InceptionV3, the
lowest-$AIC$ image model. In each case, we evaluate all subsets of
added variables and estimate random coefficients for these added variables,
in addition to those already included in the selected \textit{Review
USE} model.\footnote{We chose this approach because including all principal components
from unstructured data and all observed attributes in the candidate
set for Algorithm 1 would be too computationally burdensome.} We find that in all evaluated specifications, the variances of the
added random coefficients are zero. As a result, $AIC$ increases
and second-choice $RMSE$ does not improve. This result suggests information
is highly correlated across data types, with text providing the strongest
signal of substitution.

\paragraph{What do Principal Components Capture?}

\begin{figure}[t]
\centering{}{\small\includegraphics[width=0.8\textwidth]{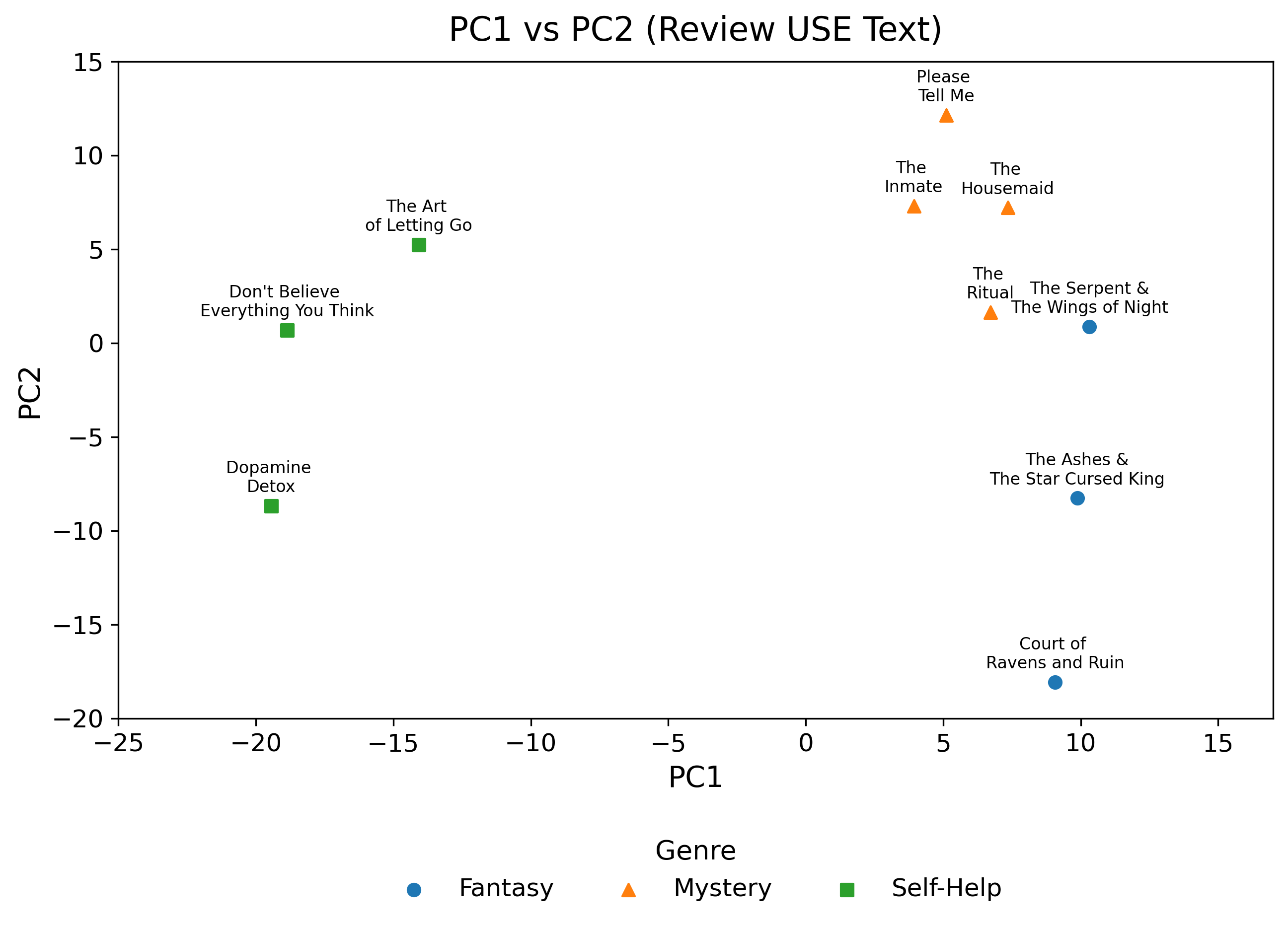}}\caption{{\small\textbf{\protect\label{fig:pc1_vs_pc2} Book locations in the
space of selected principal components (Review USE model).}}}
\end{figure}

To better understand the variation captured by embeddings, in Figure
\ref{fig:pc1_vs_pc2} we show book locations in the space of the two
principal components selected into our best-performing \textit{Review
USE} model. These principal components align with intuitive substitution
patterns: the first one (horizontal axis) separates non-fiction on
the left from fiction on the right, while the second (vertical axis)
further distinguishes science fiction from mystery. Crucially, the
variation in these principal components goes beyond separating genres.
For example, \textit{The Housemaid} and \textit{The Inmate} are by
the same author, and\textit{ The Serpent \& The Wings} and \textit{The
Ashes \& The Star Cursed King} are from the same book series. The
two principal components detect this similarity from consumers' reviews
of these books. As evident from $RMSE$ comparisons in Figure \ref{fig:model_comparison_rmse},
this additional variation helps recover substitution patterns better
and improves counterfactual predictions.

\begin{table}[!t]
\begin{centering}
\resizebox{\textwidth}{!}{
\centering
\textbf{Panel A. Predicted Second-Choice Probabilities when First Choice is \textit{Dopamine Detox (S)}}
}
\resizebox{\textwidth}{!}{
\begin{tabular}{c c c c c c c c}
\toprule
\multicolumn{2}{c}{Experimental Data} & \multicolumn{2}{c}{Plain Logit} & \multicolumn{2}{c}{Attribute-Based Mixed Logit} & \multicolumn{2}{c}{Review-Based Mixed Logit} \\
\cmidrule(lr){1-2} \cmidrule(lr){3-4} \cmidrule(lr){5-6} \cmidrule(lr){7-8}
Book & Prob. & Book & Prob. & Book & Prob. & Book & Prob. \\
\midrule
Don't Believe (S) & 0.353 & Please Tell Me (M) & 0.171 & Don't Believe (S) & 0.220 & Don't Believe (S) & 0.266 \\
Art of Letting Go (S) & 0.249 & The Inmate (M) & 0.147 & Please Tell Me (M) & 0.153 & Art of Letting Go (S) & 0.169 \\
Please Tell Me (M) & 0.112 & Don't Believe (S) & 0.143 & Art of Letting Go (S) & 0.142 & Please Tell Me (M) & 0.134 \\
The Inmate (M) & 0.094 & The Housemaid (M) & 0.139 & The Housemaid (M) & 0.134 & The Inmate (M) & 0.123 \\
The Housemaid (M) & 0.057 & Art of Letting Go (S) & 0.106 & The Inmate (M) & 0.131 & The Housemaid (M) & 0.101 \\
Serpent \& Wings (F) & 0.042 & Court of Ravens (F) & 0.096 & Court of Ravens (F) & 0.101 & Court of Ravens (F) & 0.070 \\
Court of Ravens (F) & 0.034 & Serpent \& Wings (F) & 0.088 & Serpent \& Wings (F) & 0.060 & Serpent \& Wings (F) & 0.057 \\
The Ritual (M) & 0.031 & The Ritual (M) & 0.057 & The Ritual (M) & 0.031 & The Ritual (M) & 0.043 \\
Ashes \& Star (F) & 0.030 & Ashes \& Star (F) & 0.054 & Ashes \& Star (F) & 0.027 & Ashes \& Star (F) & 0.037 \\
\bottomrule
\end{tabular}
}
\vspace{0.5cm}

\resizebox{\textwidth}{!}{
\centering
\textbf{Panel B. Predicted Second-Choice Probabilities when First Choice is \textit{Please Tell Me (M)}}
}
\resizebox{\textwidth}{!}{
\begin{tabular}{c c c c c c c c}
\toprule
\multicolumn{2}{c}{Experimental Data} & \multicolumn{2}{c}{Plain Logit} & \multicolumn{2}{c}{Attribute-Based Mixed Logit} & \multicolumn{2}{c}{Review-Based Mixed Logit} \\
\cmidrule(lr){1-2} \cmidrule(lr){3-4} \cmidrule(lr){5-6} \cmidrule(lr){7-8}
Book & Prob. & Book & Prob. & Book & Prob. & Book & Prob. \\
\midrule
The Inmate (M) & 0.325 & The Inmate (M) & 0.153 & The Inmate (M) & 0.162 & The Inmate (M) & 0.173 \\
The Housemaid (M) & 0.250 & Don't Believe (S) & 0.149 & The Housemaid (M) & 0.151 & The Housemaid (M) & 0.169 \\
Don't Believe (S) & 0.088 & The Housemaid (M) & 0.145 & Don't Believe (S) & 0.136 & Don't Believe (S) & 0.115 \\
Art of Letting Go (S) & 0.066 & Dopamine Detox (S) & 0.137 & Dopamine Detox (S) & 0.115 & Court of Ravens (F) & 0.107 \\
Serpent \& Wings (F) & 0.066 & Art of Letting Go (S) & 0.110 & Art of Letting Go (S) & 0.106 & Serpent \& Wings (F) & 0.107 \\
Dopamine Detox (S) & 0.063 & Court of Ravens (F) & 0.100 & Court of Ravens (F) & 0.102 & Dopamine Detox (S) & 0.101 \\
Court of Ravens (F) & 0.058 & Serpent \& Wings (F) & 0.092 & Serpent \& Wings (F) & 0.100 & Art of Letting Go (S) & 0.095 \\
The Ritual (M) & 0.056 & The Ritual (M) & 0.059 & The Ritual (M) & 0.064 & The Ritual (M) & 0.068 \\
Ashes \& Star (F) & 0.029 & Ashes \& Star (F) & 0.056 & Ashes \& Star (F) & 0.064 & Ashes \& Star (F) & 0.063 \\
\bottomrule
\end{tabular}
}
\vspace{0.5cm}

\resizebox{\textwidth}{!}{
\centering
\textbf{Panel C. Predicted Second-Choice Probabilities when First Choice is \textit{Ashes \& Star (F)}}
}
\resizebox{\textwidth}{!}{
\begin{tabular}{c c c c c c c c}
\toprule
\multicolumn{2}{c}{Experimental Data} & \multicolumn{2}{c}{Plain Logit} & \multicolumn{2}{c}{Attribute-Based Mixed Logit} & \multicolumn{2}{c}{Review-Based Mixed Logit} \\
\cmidrule(lr){1-2} \cmidrule(lr){3-4} \cmidrule(lr){5-6} \cmidrule(lr){7-8}
Book & Prob. & Book & Prob. & Book & Prob. & Book & Prob. \\
\midrule
Serpent \& Wings (F) & 0.275 & Please Tell Me (M) & 0.159 & Please Tell Me (M) & 0.184 & Please Tell Me (M) & 0.176 \\
Court of Ravens (F) & 0.243 & The Inmate (M) & 0.136 & The Inmate (M) & 0.158 & The Housemaid (M) & 0.152 \\
The Inmate (M) & 0.105 & Don't Believe (S) & 0.133 & The Housemaid (M) & 0.138 & The Inmate (M) & 0.150 \\
The Ritual (M) & 0.082 & The Housemaid (M) & 0.129 & Serpent \& Wings (F) & 0.123 & Court of Ravens (F) & 0.117 \\
Please Tell Me (M) & 0.080 & Dopamine Detox (S) & 0.122 & The Ritual (M) & 0.097 & Serpent \& Wings (F) & 0.104 \\
The Housemaid (M) & 0.070 & Art of Letting Go (S) & 0.098 & Court of Ravens (F) & 0.086 & Don't Believe (S) & 0.086 \\
Don't Believe (S) & 0.052 & Court of Ravens (F) & 0.089 & Don't Believe (S) & 0.081 & Dopamine Detox (S) & 0.080 \\
Art of Letting Go (S) & 0.048 & Serpent \& Wings (F) & 0.082 & Art of Letting Go (S) & 0.071 & Art of Letting Go (S) & 0.072 \\
Dopamine Detox (S) & 0.045 & The Ritual (M) & 0.053 & Dopamine Detox (S) & 0.061 & The Ritual (M) & 0.063 \\
\bottomrule
\end{tabular}
}
\vspace{0.5cm}

\par\end{centering}
\caption{{\small\textbf{\protect\label{tab:predicted_substitution}Predicted
second-choice probabilities and their data counterparts. }}{\small Letters
in parentheses indicate book genres: F=Fantasy, M=Mystery, and S=Self-Help.}}
\end{table}

\paragraph{Implied Substitution Patterns}

To illustrate how our approach improves predictions of substitution
patterns, in Table \ref{tab:predicted_substitution} we compare predicted
second-choice probabilities with their counterparts observed in the
experimental data for three of the books.

Panel A examines substitution patterns for the self-help book \textit{Dopamine
Detox}. The two other self-help books are, by far, its closest substitutes
in the data. The plain logit model misidentifies these substitutes,
incorrectly predicting that people would switch to books \textit{Please
Tell Me} and \textit{The Inmate}---two popular books with the largest
market shares. The attribute-based mixed logit correctly identifies
\textit{Don\textquoteright t Believe Everything You Think} as the
closest substitute but mispredicts the second one, likely due to its
over-reliance on estimated fixed effects. By contrast, our review-based
model is the only one that correctly predicts all five closest substitutes
in the correct order.

Panel B shows a similar example with the substitutes for the mystery
book \textit{Please Tell Me}. The plain logit model mispredicts second-choice
probabilities, incorrectly suggesting that the second-closest substitute
is a self-help book. By contrast, both the attribute-based mixed logit
and our review-based logit capture strong within-genre substitution,
correctly identifying the top three closest substitutes. Additionally,
the review-based model recognizes that \textit{The Inmate }and \textit{The
Housemaid} have significantly higher second-choice probabilities than
the third-closest substitute---an insight that the attribute-based
model misses.

These examples illustrate that, beyond reducing $RMSE$ and predicting
second-choice probabilities more accurately on average, our approach
can learn \textit{which} products are the closest substitutes.

Despite these favorable examples, our approach does not always accurately
capture substitution. Panel C shows an example where the review-based
model misidentifies the closest substitutes for the fantasy book \textit{The
Ashes \& The Star-Cursed King}. In fact, all three models fail, incorrectly
predicting mystery and self-help books as the closest substitutes.
These deviations from observed second choices suggest that there is
not enough variation in the first-choice data to reliably estimate
substitution patterns for some alternatives.

Finally, to illustrate how estimated substitution patterns affect
counterfactuals, we perform a stylized merger simulation in Appendix
\ref{sec:implications_for_mergers}. We find that the selected review-based
model predicts substantially different post-merger prices than benchmark
models.

\paragraph*{Relation Between In-Sample and Counterfactual Performance}

Although we select models based on first-choice $AIC$, our experimental
data allows us to verify whether this selection algorithm indeed chooses
specifications with the best counterfactual performance. Two findings
support our choice of $AIC$ as a model selection criterion. First,
across all specifications with principal components considered by
our model selection algorithm, the correlation between first-choice
$AIC$ and counterfactual second-choice $RMSE$ is  0.78. Second,
$AIC$ selects the specification that has the lowest second-choice
$RMSE$ across all considered specifications.

Lastly, using $BIC$ instead of $AIC$ selects the same specification,
as does a five-fold cross-validation procedure on the first choices.
This reassures us that the key results are robust to the choice of
model selection method.

\section{Application to Online Retail Data \protect\label{sec:comscore_application}}

Next, we apply our approach to choice data from several online markets.
Our first goal is to show that this approach can be applied broadly
across categories without being tailored. Our second goal is to determine
which types of unstructured data best predict substitution patterns
in various categories and to offer practical guidance on what data
researchers should collect for demand estimation.

\subsection{Data \protect\label{subsec:comscore_data}}

We use purchase data from the 2019-2020 Comscore Web Behavior Panel.
Specifically, we use the dataset constructed by \citet{greminger2023time},
who classify over 12 million unique products from Amazon.com---browsed
or purchased by Comscore panelists---into narrowly defined categories.
This dataset also matches purchases with daily product price histories
obtained from the third-party database Keepa.com.\footnote{We do not observe product rankings, so we do not include them in our
demand models.} We focus on Amazon.com due to the large volume of Amazon purchases
in the Comscore dataset.

\begin{figure}[t]
\begin{centering}
\includegraphics[width=14cm]{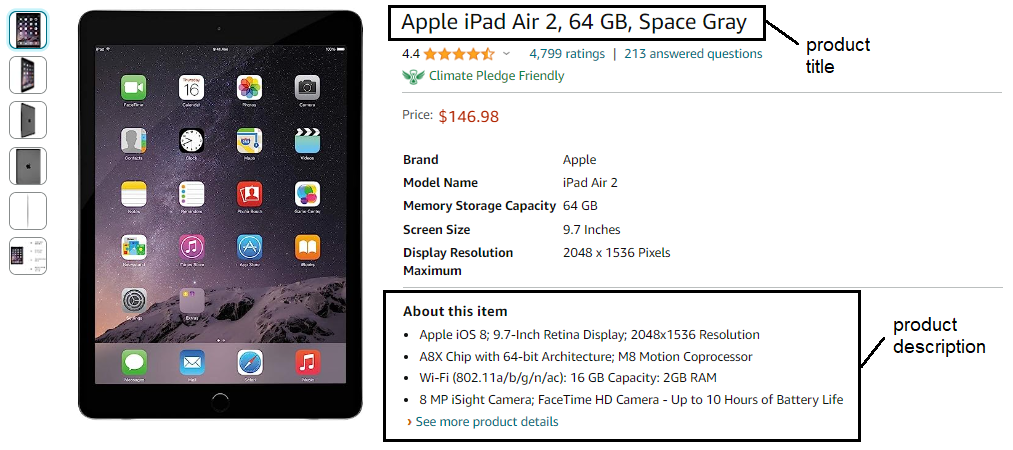}
\par\end{centering}
\begin{centering}
\includegraphics[width=14cm]{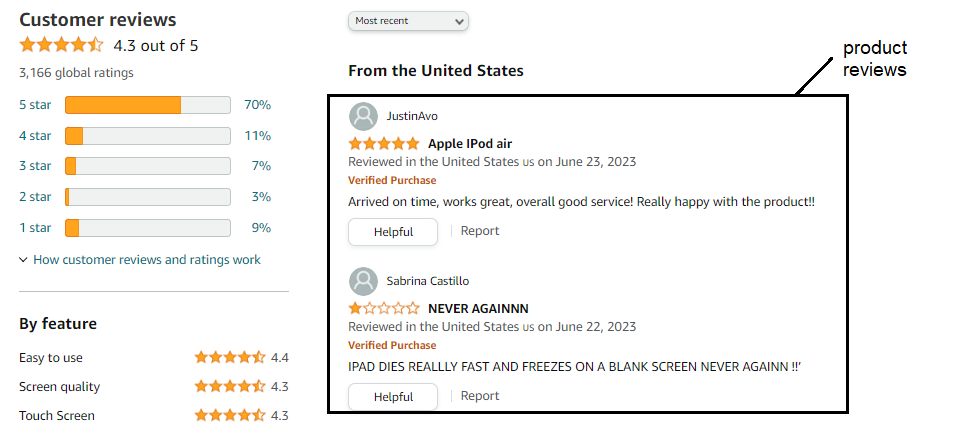}
\par\end{centering}
\caption{{\small\protect\label{fig:=000020example_screenshot}}{\small\textbf{Example:
image and text data collected from product detail pages.}}}
\end{figure}

We combine purchase data with image and text data we collected from
Amazon product detail pages (Figure \ref{fig:=000020example_screenshot}).
We use the default product photo for image embeddings. For text embeddings,
we use product titles, product descriptions (i.e., bullet points describing
the product), and 100 most recent reviews for each product. In estimation,
we treat text and image embeddings as fixed over time. While sellers
could theoretically modify images and textual descriptions to reposition
their products, we find that product images, titles, and descriptions
very rarely change over time (see Appendix \ref{sec:changes_in_embeddings}).

We apply our method to 40 product categories spanning clothing (shirts,
blouses, underwear, sleepwear), household goods (paper towels, trash
bags, batteries), office supplies (pens, markers, printing paper),
groceries (tea, coffee, bottled water), pet food (wet food, treats),
electronics (tablets, monitors, headphones, memory cards, media players),
and video games (PC, Nintendo, Xbox, PlayStation). We select the 15
most-purchased products in each category and retain only those categories
where these products collectively account for at least 2,000 purchases.
These selection criteria ensure that we can estimate product fixed
effects precisely, which helps capture time-invariant differences
across products, and that we have sufficient variation to estimate
substitution patterns.

For four electronics categories, we collect standard attributes from
product detail pages to compare our approach with attribute-based
mixed logit: 18 attributes for ``Tablets,'' 13 for ``Monitors,''
20 for ``Memory Cards,'' and 14 for ``Headphones'' (see Appendix
\ref{sec:Attributes-for-Electronic} for a full list).

\subsection{Estimation Results \protect\label{subsec:comscore_results}}

We first compare our approach to benchmark models in the four electronics
categories for which we collected attribute data. Since we have a
relatively large number of attributes in these categories, we reduce
their dimensionality via PCA and apply Algorithm 1 for model selection.\footnote{We normalize each attribute to have mean zero and variance one before
applying PCA to prevent attributes with larger variance or units from
dominating the principal components.} As in   Section \ref{sec:choice_experiment}, we let the candidate
set  include  price and the first $P=6$ principal components.

Table \ref{tab:aic_comparison} shows the results. In all four categories,
unstructured data meaningfully improves model fit relative to both
plain logit and  attribute-based mixed logit, suggesting that our
approach captures information about substitution patterns beyond that
contained in standard attributes. This finding is particularly noteworthy
given that one might expect technical specifications of electronics
products---captured by our extensive list of attributes---to be
highly relevant for consumer choices.

\begin{table}
\begin{centering}
\begin{centering}
\begin{tabular}{>{\raggedright\arraybackslash}p{6cm}>{\centering\arraybackslash}p{2cm}>{\centering\arraybackslash}p{3cm}}
\hline
& {\footnotesize{$AIC$}} & {\footnotesize{$\Delta AIC$}\textbf{ Relative to Plain Logit}}\tabularnewline
\hline
{\footnotesize\textbf{Category: Tablets}} &  & \tabularnewline
{\footnotesize Mixed Logit with Attributes} & {\footnotesize 4293.2} & {\footnotesize -54.9}\tabularnewline
{\footnotesize Mixed Logit with Unstructured Data} & {\footnotesize 4236.6} & {\footnotesize -111.5}\tabularnewline

{\footnotesize\textbf{Category: Monitors}} &  & \tabularnewline
{\footnotesize Mixed Logit with Attributes} & {\footnotesize 1376.4} & {\footnotesize -12.3}\tabularnewline
{\footnotesize Mixed Logit with Unstructured Data} & {\footnotesize 1366.4} & {\footnotesize -22.4}\tabularnewline

{\footnotesize\textbf{Category: Memory Cards}} &  & \tabularnewline
{\footnotesize Mixed Logit with Attributes} & {\footnotesize 2709.3} & {\footnotesize -72.6}\tabularnewline
{\footnotesize Mixed Logit with Unstructured Data} & {\footnotesize 2685.1} & {\footnotesize -96.8}\tabularnewline

{\footnotesize\textbf{Category: Headphones}} &  & \tabularnewline
{\footnotesize Mixed Logit with Attributes} & {\footnotesize 9882.0} & {\footnotesize -4.5}\tabularnewline
{\footnotesize Mixed Logit with Unstructured Data} & {\footnotesize 9875.8} & {\footnotesize -10.7}\tabularnewline

\hline
\end{tabular}
\par\end{centering}
\par\end{centering}
\caption{{\small\textbf{\protect\label{tab:aic_comparison}$AIC$ comparison
of our approach and attribute-based mixed logit.}}{\small{} The table
reports the $AIC$ of the specification selected by Algorithm 1 within
each class of demand models.}}
\end{table}

These performance differences translate into different substitution
patterns. For example, consider diversion ratios for the category
of tablets (see Appendix Tables \ref{tab:diversions_logit}-\ref{tab:diversions_unstructured}).
 Relative to the plain logit, the selected \textit{Description ST}
model produces more intuitive substitution patterns: kids\textquoteright{}
tablets are close substitutes (Fire Kids and Fire HD8 Kids), iPads
are close substitutes (iPad 9.7 and iPad 10.2), and so on. The predicted
diversions in this model also differ markedly from those in the mixed
logit with attributes. However, in the absence of second choice data,
we cannot directly test how well different models predict diversions
counterfactually, as we do in Section \ref{sec:choice_experiment}.
We therefore rely on the fit comparisons reported in Table \ref{tab:aic_comparison},
which favor the \textit{Description ST} model.

Next, we expand the analysis to all 40 categories. Because observed
attributes are not available in all of them, we benchmark results
against the plain logit model (see Appendix Table \ref{tab:category_level_results}
and Appendix Figure \ref{fig:aic_improvements}). As a rule of thumb,
we consider a model to have strong support over a simpler model if
it lowers $AIC$ by at least $2.0$, and very strong support if it
lowers $AIC$ by at least $5.0$.\footnote{When a simpler model is nested within a more complex one with only
one additional parameter, applying these $AIC$ thresholds is approximately
equivalent to conducting a likelihood ratio test at the 5\% and 1\%
significance levels. In this case, the reduction in $AIC$ is given
by $\Delta AIC=2-\lambda_{LR}$, where $\lambda_{LR}$ is the likelihood
ratio statistic, meaning the $\Delta AIC$ thresholds of $2.0$ and
$5.0$ correspond to the likelihood ratio thresholds $4.0$ and $7.0$
(p-values $0.0455$ and $0.008$).} We find that our approach reduces \textit{$AIC$} by at least $5.6$
in every category. On average, the reduction is 23.3, reaching 111.5
in some categories.\footnote{In some categories, our estimates of the price coefficient $\alpha$
are positive, likely due to correlation between prices and unobserved
demand shocks, even after accounting for product fixed effects. While
instrumental variables could address this, we do not pursue this approach,
as addressing price endogeneity across many product categories is
orthogonal to the contribution of our paper. Instead, we focus on
counterfactuals, such as the diversion ratios from removing a product,
which remain valid even if the price coefficient is positive.} These results suggest that our approach consistently captures signals
of substitution from unstructured data across a wide range of categories.

\begin{figure}
\centering{}\includegraphics[width=0.9\textwidth]{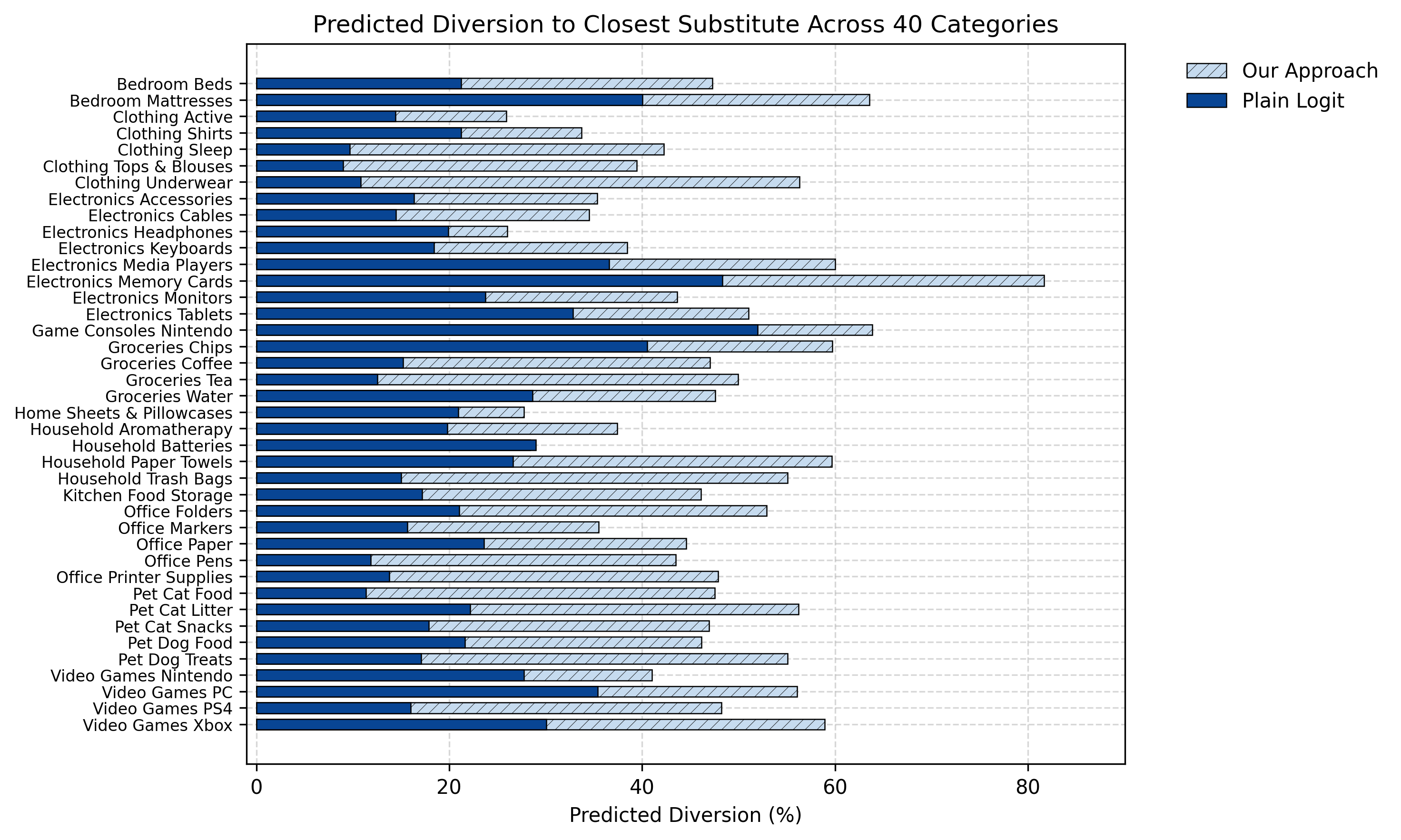}\caption{{\small\textbf{\protect\label{fig:diversions_estimated}Estimated
diversion ratios to closest substitutes. }}{\small This figure plots
the estimated diversion ratios to the closest substitutes, $\max_{k}\ensuremath{\hat{s}_{j\rightarrow k}}$,
averaged across products $j$ in each category.}}
\end{figure}

A well-known limitation of the plain logit is that diversion ratios
depend only on market shares rather than attribute similarity, leading
to overly flat diversions \citep{conlon2023estimating}. Figure \ref{fig:diversions_estimated}
plots the estimated diversion ratio to the closest substitute, $\max_{k}\ensuremath{\hat{s}_{j\rightarrow k}}$,
averaged across products $j$. The plain logit yields relatively small
diversions, about 22\% on average, while our approach increases this
average to 47\% and, in some categories, to as high as 60-80\%. These
results indicate that our approach produces substantially more variable
diversion ratios, suggesting it better identifies close substitutes.

\subsection{Relevance of Different Data Types}

Lastly, we analyze which types of unstructured data yield the largest
fit improvements. It would be misleading to report only the best-performing
model in each product category, as multiple text or image models may
achieve similar $AIC$ improvements over plain logit. We therefore
follow the statistical literature on model selection and report Akaike
weights \citep{burnham2004multimodel}. Formally, let $d$ denote
data type (images, titles, descriptions, or reviews), and let $AIC_{d}$
denote the lowest $AIC$ among all specifications that use data type
$d$.\footnote{The summation in the denominator is over all four data types.}
 For each data type $d$, we compute the Akaike weight $w_{d}$ as
\begin{equation}
w_{d}=\frac{\exp(-\Delta_{d}/2)}{\sum_{k}\exp(-\Delta_{k}/2)}\label{eq:akaike_weight}
\end{equation}
where $\Delta_{d}=AIC_{d}-AIC_{min}$  and $AIC_{min}=\min_{d}AIC_{d}$
is the lowest $AIC$ across \textit{all} four data types. We interpret
$w_{d}$ heuristically as the posterior probability that the mixed
logit model based on data type $d$ is the best model given the data.\footnote{More precisely, in large samples, $w_{d}$ reflects the probability
that this class of models is, in fact, the best model for the data
in the sense of Kullback-Leibler information \citep[p.272]{burnham2004multimodel}.}

\begin{figure}
\centering{}\includegraphics[width=0.9\textwidth]{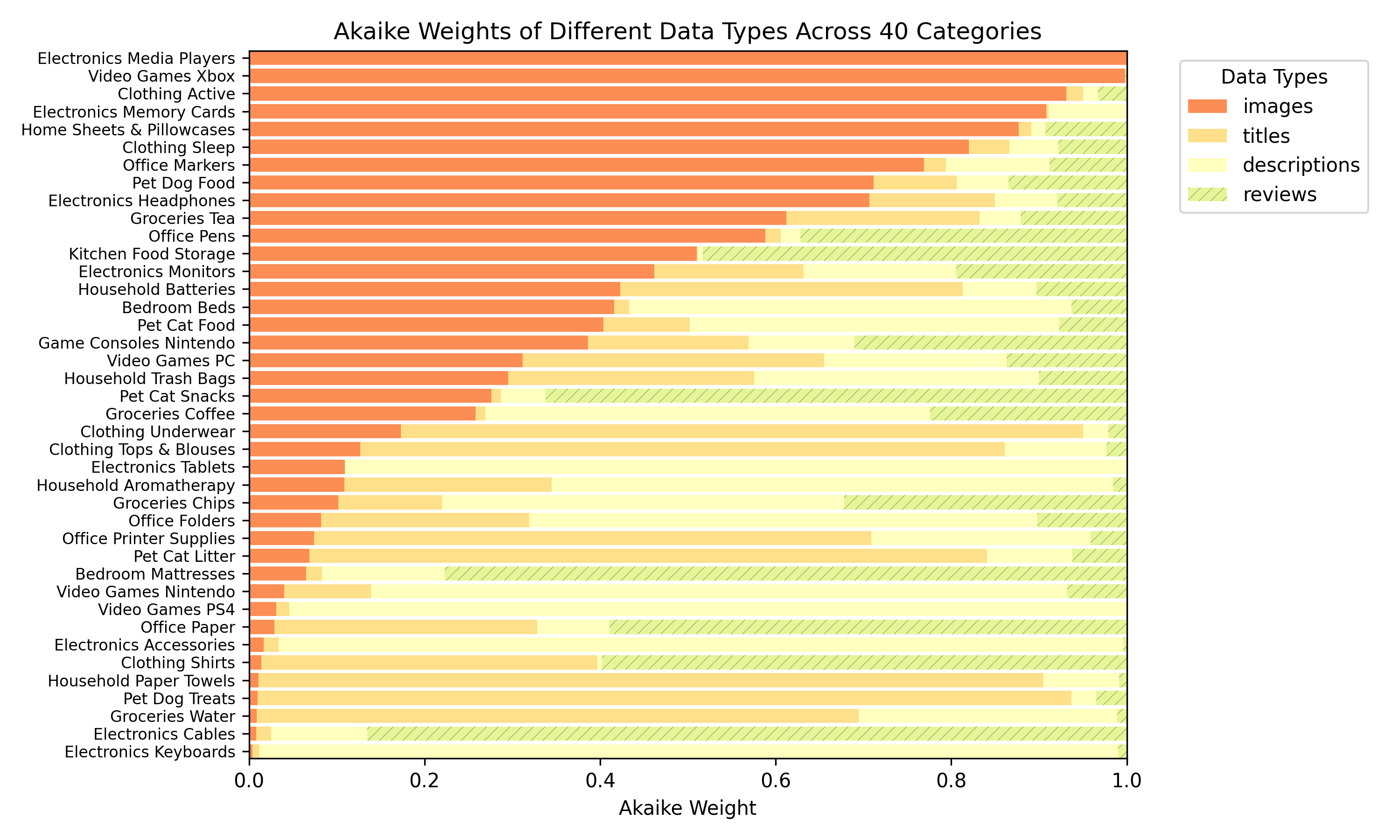}\caption{\textbf{\protect\label{fig:akaike_weights}}{\small\textbf{Akaike
weights of different types of unstructured data across 40 product
categories. }}{\small These weights reflect the relative  performance
of data types  in capturing substitution patterns  from choice data.
We compute Akaike weights using the formula in (\ref{eq:akaike_weight}).}}
\end{figure}

Figure \ref{fig:akaike_weights} shows the estimated Akaike weights
by category and data type, revealing substantial variation across
categories. Many results are difficult to anticipate \textit{a priori}.
For example, although one might expect visual features to be most
relevant for clothing, only two of the five clothing categories (``Activewear''
and ``Sleep'') place more weight  on images than text. In ``Tops
\& Blouses'' and ``Underwear,'' product titles are most predictive
of substitution, while in ``Shirts,'' reviews perform best. Similarly,
we may expect descriptions and reviews of video games to be most informative,
analogous to books. In practice, we find that images contain substantially
more information about substitution in the ``Video Games Xbox''
category.

These results highlight that researchers cannot reliably predict which
data types best capture substitution. In practice, it is therefore
important to collect multiple data types and  use model selection
to choose the  one that best captures substitution.

\section{A Practitioner's Guide for Using Text and Images}

In this section, we provide guidance for practitioners interested
in applying our method. We highlight several practical lessons from
Sections \ref{sec:choice_experiment} and \ref{sec:comscore_application},
clarify which counterfactuals our approach can and cannot accommodate,
and suggest promising directions for future research.

\subsection{Data Choice and Model Selection\protect\label{subsec:practitioners_guide}}

The main takeaway from our results is that text and image data contain
valuable information about substitution patterns, making these unstructured
data useful for demand estimation. At the same time, our multi-category
analysis in Section \ref{sec:comscore_application} shows that researchers
may not be able to predict in advance which data type will best capture
substitution in a given category. We therefore recommend collecting
various types of unstructured data and performing model selection
as in Sections \ref{sec:choice_experiment} and \ref{sec:comscore_application}.
Algorithm 1 in Section \ref{sec:proposed_approach} provides a practical
heuristic for model selection: in our experiment, it successfully
identifies the specification that produces the best counterfactual
second-choice predictions.

A natural question is whether researchers should add observed attributes
to our approach when they are available. In our application, adding
observed attributes to the selected review-based model does not improve
fit (see Section \ref{subsec:second_choice_validation}). One explanation
is that unstructured data may already encode the choice-relevant attributes,
rendering structured attributes redundant. Nevertheless, in other
applications, researchers may still want to test whether including
observed attributes helps estimate substitution patterns. For example,
researchers can expand the set of candidate variables in Algorithm
1 to include the observed attributes themselves or, when many attributes
are available, their principal components.

\subsection{Counterfactual Analysis and its Boundaries \protect\label{subsec:counterfactuals}}

Our approach is well suited for applications that assess consumer
responses to price changes while holding embeddings fixed, including
optimal pricing, merger simulations, corrective taxes, and markup
estimation \citep{berry2024nonparametric}. In these applications,
embeddings may be correlated with prices in the observed data, but
they, and hence the demand system, should be held fixed in counterfactual
simulations. Under this assumption, researchers can compute counterfactual
market shares, prices, and welfare using standard tools from demand
analysis. For instance, consumer welfare can be computed as the area
under the demand curve \citep{small1981applied}. The fact that embeddings
have no direct interpretation does not preclude these analyses.

Our approach can also predict responses to changes in product availability,
for example, to study firms' assortment choices or consumer surplus
from new products \citep{petrin2002quantifying}. If new products
launches are observed in the data, researchers can apply our method
directly, since our results show it performs well at predicting consumer
responses to product removals. On the other hand, when evaluating
a hypothetical new product, researchers must construct its embeddings
from pre-launch information, such as images and descriptions from
marketing materials or early retailer listings.

By contrast, our approach is less suitable in applications where embeddings
may change in counterfactuals. For example, if customer reviews mention
value relative to price (e.g., \textit{``Great quality for the price!''}
or \textit{``Good product, but overpriced.''}), embeddings constructed
from them may shift when prices change. Researchers can diagnose this
issue by checking how frequently price-related language appears in
reviews and, if necessary, modeling how price changes affect embeddings.
Another example is simulations of mergers that may induce firms to
redesign products, which would alter embeddings and hence substitution
patterns. Studying such mergers is not possible with our approach
unless researchers impose additional assumptions on firm conduct and
on how embeddings map into attributes that firms can control.

\subsection{Future Research Directions}

Our paper shows that text and image data contain valuable and easily
extractable information for estimating substitution patterns. This
finding opens several promising directions for future research.

First, newer ML models might extract substitution patterns from texts
and images more effectively. For example, Qwen3, OpenAI, and Gemini
text embeddings perform robustly well across diverse natural language
tasks and may thus generalize to demand estimation \citep{lee2025gemini,zhang2025qwen3}.
Because our method is modular, researchers can easily try alternative
embeddings and test whether that improves performance. In addition,
researchers can fine-tune existing models to construct embeddings
optimized for counterfactual predictions. An open question is how
much fine-tuning improves upon pre-trained models and whether these
gains come without excessive computational costs. To facilitate future
research on alternative embedding models, we make our experimental
dataset and estimation code publicly available.\footnote{Replication codes and experimental data are available in our public
repository: \href{https://github.com/ilyamorozov/DeepLogitReplication}{github.com/ilyamorozov/DeepLogitReplication}.
Separately, the Python package for implementing our method is available
on PyPI and at: \href{https://github.com/deep-logit-demand/deeplogit}{github.com/deep-logit-demand/deeplogit}.}

Second, our method incorporates embeddings into the demand model
as if they were the true attributes driving substitution, when in
practice embeddings may be only imperfect proxies of these attributes.
Our results show that, despite this limitation, embeddings are sufficiently
informative to improve counterfactual predictions. Nevertheless, future
research should explore whether explicitly accounting for error in
these proxies can further improve performance. For an early contribution
in this area, see \citet{christensen2026unstructured}.

Third, other demand models may leverage unstructured data more effectively.
In an earlier version of this paper, we tested the pairwise combinatorial
logit model of \citet{Koppelman_Wen_2000}, which allows pairwise
utility correlations to depend on distances in the embedding space.
That model performed worse than mixed logit, not only when using unstructured
data but also when using distances in observed attributes,  suggesting
that functional form assumptions can significantly impact counterfactual
performance. A fruitful direction for future research might be to
combine unstructured data with more flexible demand models, such as
that in \citet{compiani2022market}.

Finally, because our focus is on recovering substitution patterns,
we largely abstract away from price endogeneity. Endogeneity is not
a concern in our experiment because we randomize prices. In the e-commerce
application, we rely on product fixed effects to capture unobserved
attributes (e.g., quality) that may correlate with prices. More generally,
prices may correlate with unobserved demand shocks that vary across
markets and over time. To address this, one could combine our approach
with existing methods for handling price endogeneity using instrumental
variables \citep{goolsbee2004consumer,BLP_micro_2004}.\footnote{For example, one could estimate product-market fixed effects and random
coefficient variances via maximum likelihood, and then estimate a
two-stage least squares regression of the estimated fixed effects
on prices and principal components. The principal components of competing
products could serve as excluded instruments for prices.}

\section{Conclusion \protect\label{sec:concluding_remarks}}

In this paper, we show how researchers can incorporate unstructured
text and image data in demand estimation to recover substitution patterns.
Our approach extracts low-dimensional features from product images
and textual descriptions, integrating them into a standard mixed logit
model with random coefficients. Using experimental data, we show that
our approach outperforms standard attribute-based models in counterfactual
predictions of second choices. We further validate our method with
e-commerce data across dozens of categories and find that text and
image data consistently help identify close substitutes within each
category.

\clearpage{}

\begin{onehalfspace}
\begin{spacing}{1}

\bibliographystyle{ecta}
\bibliography{references}

\end{spacing}
\end{onehalfspace}

\clearpage{}

\appendix
\renewcommand{\thetable}{\Alph{appendix}\arabic{table}} 
\renewcommand{\thefigure}{\Alph{appendix}\arabic{figure}}
\setcounter{appendix}{1} 
\setcounter{table}{0} 
\setcounter{figure}{0}

\section*{Online Appendix}

\section{Text Processing Steps \protect\label{sec:appx_text_preprocessing}}

Before computing embeddings, we pre-process text as follows. We keep
product titles unchanged because they typically consist of only 10-15
words. For product descriptions, we merge the text from all bullet
points and apply the embedding models to the merged text. For customer
reviews, we transform the text of each review into a vector of word
occurrences or an embedding, and we average these vectors or embeddings
across all reviews of a given product. 

For both bag-of-words models, we further pre-process text data by
removing stopwords and lemmatizing words. Stopwords are removed using
the standard dictionary of common English words in the NLTK package.
We lemmatize words using the WordNet Lemmatizer from the same package,
NLTK. We then convert each processed text into a vector of word occurrences
(or weighted word occurrences for TF-IDF). By contrast, the USE and
ST models have built-in text pre-processing, so we apply them directly
to the raw text.

\section{Book Selection Procedure \protect\label{sec:appendix_book_selection}}

The set of books shown to participants can significantly affect the
performance of demand models. Our goal was to select books in a way
that does not favor any particular specification. To this end, we
maximized variation in both observed attributes as well as in text-
and image-based embeddings.

We chose books from three genres---Mystery, Fantasy, and Self-Help---and
identified 20 books of each genre from Amazon's bestseller lists.
We included books from three different genres in order to generate
meaningful variation in this structured attribute, which we anticipated
would be predictive of substitution patterns.

To avoid arbitrary book selection within each genre, we implemented
the following algorithm. We considered all possible combinations of
ten books such that the three genres are roughly equally represented.
Among these, we choose the combination of books that maximized the
variance of text and image embeddings.\footnote{We computed the variance of image embeddings (using the VGG19 model)
and the variance of text embeddings (using the USE model), then averaged
the two. We performed a brute-force search over all possible sets
of ten books and selected the set that maximized this average variance.} The final set, shown in Figure \ref{fig:book_covers}, includes four
mystery books, three fantasy books, and three self-help books.

\section{Choice Survey: Sanity Checks \protect\label{sec:sanity_checks}}

\label{app_sec:=000020Choice=000020Survey=000020Validity}

Because choices in our experiment were not incentivized, it is crucial
to verify that participants made meaningful choices. Several summary
statistics suggest that participants took the choice tasks seriously.
First, only 50 of the initially recruited participants (less than
1\%) completed the entire study in under one minute and thus were
excluded from the sample. The remaining participants spent, on average,
7 minutes on the survey overall and 1.3 minutes on the choice tasks,
indicating they took their time to make careful selections and did
not mindlessly click through the survey (Figure \ref{fig:study_time_spent}).

Second, participants were disproportionately more likely to select
books of the genre they reported to be their favorite in a questionnaire
before the choice tasks (Figure \ref{fig:genres_self_reported}).
Further, their choices were consistent across the two choice tasks
(Figure \ref{fig:genres_first_second_choices}). For example, over
60\% of participants who selected a mystery book in the first task
chose another mystery book in the second choice task. These observations
suggest that participants considered their genre preferences when
making choices.

\clearpage{}

\section{Computing Second-Choice Diversion Ratios \protect\label{sec:rmse_computations}}

To compute $RMSE$ in (\ref{eq:rmse_equation}), we need to calculate
predicted second-choice diversion ratios $\ensuremath{\hat{s}_{j\rightarrow k}}$
and their data counterparts $\ensuremath{s_{j\rightarrow k}}$. Recall
that both $\ensuremath{s_{j\rightarrow k}}$ and $\ensuremath{\hat{s}_{j\rightarrow k}}$
reflect the probability that the consumer chooses book k in the second
choice task conditional on having chosen book j in the first choice
task. Using the analogy principle, we compute empirical diversions
$\ensuremath{s_{j\rightarrow k}}$ using a simple frequency estimator:

\begin{equation}
\ensuremath{s_{j\rightarrow k}}=\frac{\sum_{i=1}^{N}\mathbf{1}\{y_{i}^{(2)}=k,y_{i}^{(1)}=j\}}{\sum_{i=1}^{N}\mathbf{1}\{y_{i}^{(1)}=j\}}\quad\text{for}\quad j\neq k,\label{eq:empirical_rmse}
\end{equation}
where $y_{i}^{(1)}$ and $y_{i}^{(2)}$ are participant $i$'s first
and second choices. By contrast, to compute diversions $\ensuremath{\hat{s}_{j\rightarrow k}}$
predicted by a given demand model, we use the following identity from
\citet{conlon2021empirical}:

\begin{equation}
Pr(y_{i}^{(2)}=k|y_{i}^{(1)}=j;w_{i})=\frac{s_{k}^{(j)}(w_{i})-s_{k}(w_{i})}{s_{j}(w_{i})},\label{eq:useful_identify}
\end{equation}
where $s_{j}(w_{i})$ and $s_{k}(w_{i})$ are the unconditional choice
probabilities of books $j$ and $k$, $s_{k}^{(j)}(w_{i})$ is the
probability of choosing book $k$ after book $j$ is removed from
the choice set, and $w_{i}$ is a vector of prices $(\text{price}_{i1},\dots,\text{price}_{iJ})$
and rankings $(\text{rank}_{i1},\dots,\text{rank}_{iJ})$ that participant
$i$ encounters in the experiment. We average diversions in (\ref{eq:useful_identify})
across all participants to compute $\ensuremath{\hat{s}_{j\rightarrow k}}$:

\begin{equation}
\ensuremath{\hat{s}_{j\rightarrow k}=\frac{1}{N}\sum_{i=1}^{N}\frac{s_{k}^{(j)}(w_{i})-s_{k}(w_{i})}{s_{j}(w_{i})}}.\label{eq:predicted_rmse}
\end{equation}

\section{Implications for Pricing: Merger Simulations \protect\label{sec:implications_for_mergers}}

To illustrate how estimated substitution patterns can influence counterfactuals
of interest, we conduct a stylized simulation of horizontal mergers---a
natural application of our approach. Antitrust agencies routinely
use demand models to assess whether a hypothetical merger would lead
to a significant price increase \citep{FTC2022mergers}. If the merging
firms' products are close substitutes, the merger creates strong upward
pricing pressure, as the firm can \textquotedblleft recapture\textquotedblright{}
some consumers after raising prices. Therefore, predicting how a merged
firm would set prices requires accurate estimates of substitution
patterns.

\begin{figure}[t]
\centering{}{\small\includegraphics[width=0.8\textwidth]{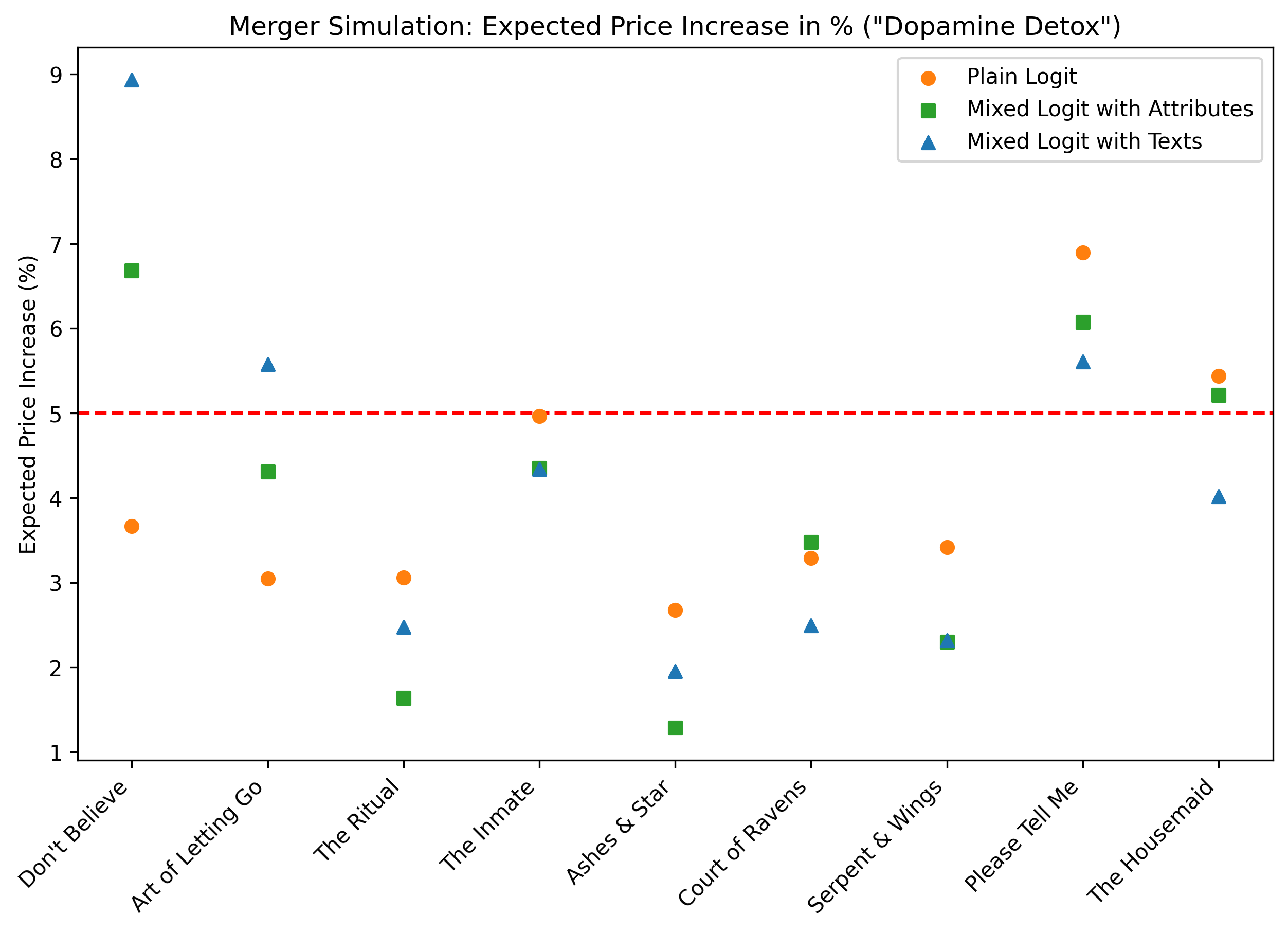}}\caption{{\small\textbf{\protect\label{fig:merger_simulations}Results of hypothetical
merger simulations. }}{\small For each simulated merger of }{\small\textit{Dopamine
Detox}}{\small{} with one other book (horizontal axis), the figure shows
the predicted average price increase among the two merging books.
The horizontal dashed line represents a hypothetical policy where
the decision-maker challenges all mergers expected to raise prices
by at least 5\%. }}
\end{figure}

We perform simulations using our demand estimates from Section \ref{sec:choice_experiment}.
For each pair of books, we compute their prices under two scenarios:
(a) when the books are owned by separate publishers competing in a
Bertrand-Nash equilibrium, and (b) when both books are owned by the
same publisher setting prices by solving a joint first-order condition.
In both cases, publishers take the prices of the other eight books
as given, fixed at \$5, and have zero marginal costs.\footnote{We fix the prices of the other books at \$5, the average value in
the experimental dataset. We do not optimize prices of other books
because the model does not include an outside option. Consumer choices
remain unchanged if all prices increase by the same amount, thus leading
to multiple equilibria.}

We recognize that restricting the choice set to ten books and excluding
an outside option makes the setting somewhat artificial. Still, we
view it as a useful way to evaluate how estimated substitution patterns
translate into counterfactuals of interest.

Figure \ref{fig:merger_simulations} illustrates the predicted price
increase resulting from the joint ownership of \textit{Dopamine Detox}
with each of the other books. We select \textit{Dopamine Detox }as
an example where our approach outperforms alternative models in capturing
substitution patterns (see Table \ref{tab:predicted_substitution}).
For each simulated merger, Figure \ref{fig:merger_simulations} reports
the average price increase across the two merging books.

As discussed in Section \ref{sec:choice_experiment}, the plain logit
model fails to capture strong within-genre substitution. Consequently,
it incorrectly identifies as the closest substitutes for \textit{Dopamine
Detox} the three mystery books, rather than other self-help ones.
This misclassification affects the predicted price changes in merger
simulations. Specifically, the plain logit model overstates the price
increase when \textit{Dopamine Detox} is merged with the mystery book
\textit{Please Tell Me} or \textit{The Housemaid}, while understating
the price increase when merged with its true closest substitutes,
\textit{The Art of Letting Go} or \textit{Don\textquoteright t Believe
Everything You Think}. These discrepancies are substantial: for within-genre
mergers, the plain logit predicts modest price increases of 3\% and
3.7\%, whereas our review-based model estimates them to be 5.6\% and
8.9\%---approximately 2 to 2.5 times higher.

The attribute-based mixed logit model produces predictions more aligned
with the review-based model. In most cases, both models deviate from
the plain logit in the same direction, yielding similar or even identical
predicted price increases. However, notable discrepancies remain---for
instance, in the case of the self-help books \textit{The Art of Letting
Go} or \textit{Don\textquoteright t Believe Everything You Think},
the attribute-based model underestimates the price increase by approximately
25\% compared to the review-based model.

These divergent predictions could lead decision-makers to different
conclusions. As an example, consider an antitrust agency that applies
a heuristic rule, challenging all mergers expected to increase prices
by more than 5\% \citep{bhattacharya2023merger}. Compared to decisions
made using the review-based model, which best captures substitution
patterns, a decision-maker relying on the plain logit model would
approve two mergers that should be challenged (\textit{Don\textquoteright t
Believe Everything You Think, The Art of Letting Go}) and challenge
one that should be approved (\textit{The Housemaid}). Similarly, a
decision-maker using the attribute-based mixed logit model would approve
a merger that should be challenged (\textit{The Art of Letting Go})
and challenge one that should be approved (\textit{The Housemaid}).

\section{Attributes Collected for Electronics Products\protect\label{sec:Attributes-for-Electronic}}

For each of the four electronics categories described in Section \ref{subsec:comscore_data},
we collect standard attributes from Amazon product detail pages. Specifically,
we include all specifications listed in the product descriptions and,
when available, those in the ``Technical Details'' section of the
product detail page. Below is the complete list of all attributes
for each category:
\begin{enumerate}
\item \textbf{Tablets}: brand, model, memory, RAM, processor speed, number
of cores, screen size, maximum resolution, charging time, battery
life, front camera, back camera, front camera megapixels, back camera
megapixels, warranty, whether the product comes with a case, and whether
it includes an Amazon Kids subscription.
\item \textbf{Monitors}: brand, screen size, maximum resolution, refresh
rate, blue light filter, frameless design, tilt adjustment, height
adjustment, flicker-free display, built-in speaker, wall-mountable
option, curved screen, and adaptive sync.
\item \textbf{Memory Cards}: brand, pack size, micro card, flash memory
type, capacity, read speed, write speed, speed class, UHS speed class,
device compatibility (smartphone, computer, camera, laptop, tablet),
X-ray proof, temperature proof, waterproof, shockproof, magnetic proof,
and whether an adapter is included.
\item \textbf{Headphones}: brand, color, connectivity, number of eartip
sets, battery life on a single charge, battery life with charging
case, deep bass feature, tangle-free design, waterproof, sweat-proof,
IPX rating, whether the product comes with a case, a microphone, and
a noise reduction option.
\end{enumerate}

\section{Changes in Text and Images Over Time\protect\label{sec:changes_in_embeddings}}

Throughout the paper, we treat text and image embeddings, as well
as the principal components extracted from them, as being fixed over
time for a given product. To validate this assumption, we examine
whether text and images change over time. 

Since we lack data on these changes for 2019-2020, we construct a
separate sample by repeatedly collecting unstructured data from Amazon's
product detail pages daily from January 23 to March 4, 2025. To keep
data collection manageable, we do not gather customer reviews and
select a subset of 11 out of 40 categories, ensuring they cover all
of Amazon's departments (e.g., ``Clothing,'' ``Food,'' ``Electronics'')
represented in the full dataset.\footnote{Recall that we average over embeddings extracted from the 100 most
recent reviews. Even though customer reviews accumulate over time
as consumers continue to write them, the average embeddings extracted
from these reviews may remain approximately constant if consumers
consistently discuss the same attributes. While we do not have review
data to verify this claim, future research should examine whether
review embeddings are stable in online markets.} The selected categories are: \textquotedblleft Shirts,\textquotedblright{}
``Coffee,\textquotedblright{} \textquotedblleft Aromatherapy,\textquotedblright{}
\textquotedblleft Mattresses,\textquotedblright{} \textquotedblleft Markers,\textquotedblright{}
\textquotedblleft Pet Litters,\textquotedblright{} \textquotedblleft Nintendo
Games,\textquotedblright{} \textquotedblleft Tablets,\textquotedblright{}
\textquotedblleft Memory Cards,\textquotedblright{} \textquotedblleft Monitors,\textquotedblright{}
and \textquotedblleft Earbuds.\textquotedblright{}  In each category,
we collect data for the products used in estimation in Section \ref{sec:comscore_application}
that were not discontinued, totaling 136 products. 

We find that product images do not change over time. Titles change
for only six products (4\%), mostly by adding or removing specific
attributes or functional benefits.  Similarly, descriptions change
for just 21 products (15\%). Most changes do not affect which product
features are revealed, but they alter which ones are immediately visible
versus being revealed only after clicking on ``See more product details.''
 Thus, we conclude that changes in text and images are not a significant
concern in our empirical application. 
\begin{center}
\clearpage{}
\begin{figure}[!t]
\begin{centering}
\includegraphics[scale=0.6]{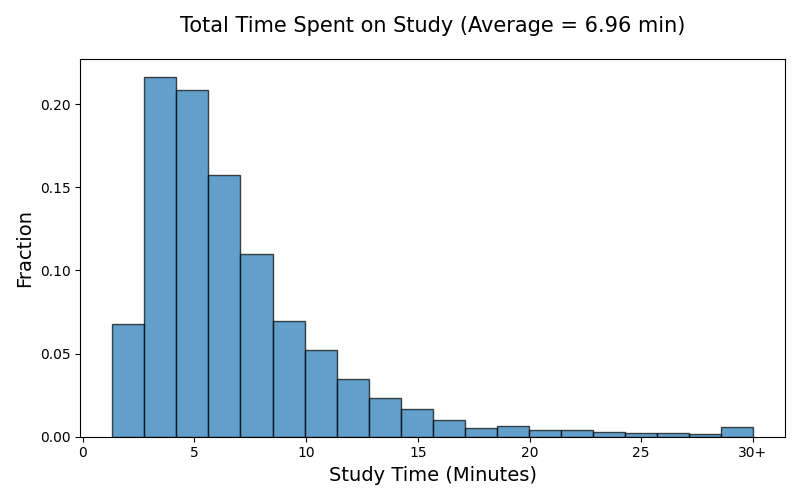}
\par\end{centering}
\bigskip{}

\begin{centering}
\includegraphics[scale=0.6]{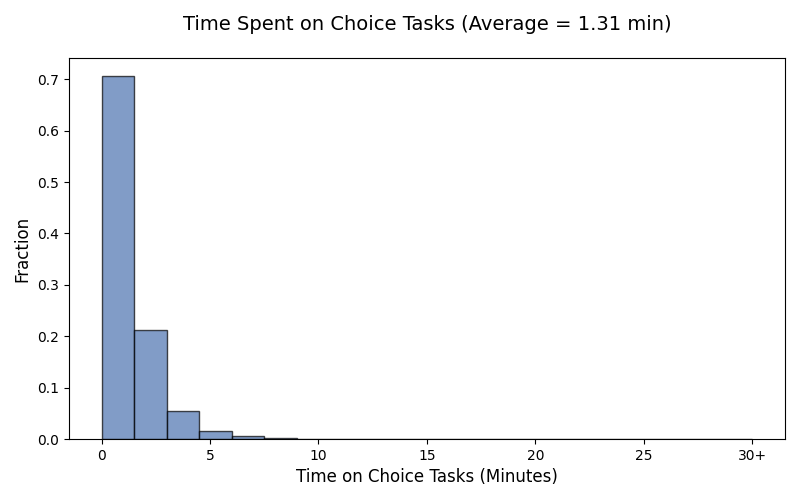}
\par\end{centering}
\caption{\textbf{\protect\label{fig:study_time_spent}}{\small\textbf{Time
spent by participants on choice tasks in the experiment.}}}
\end{figure}
\par\end{center}

\begin{center}
\begin{figure}[t]
\begin{centering}
\includegraphics[scale=0.42]{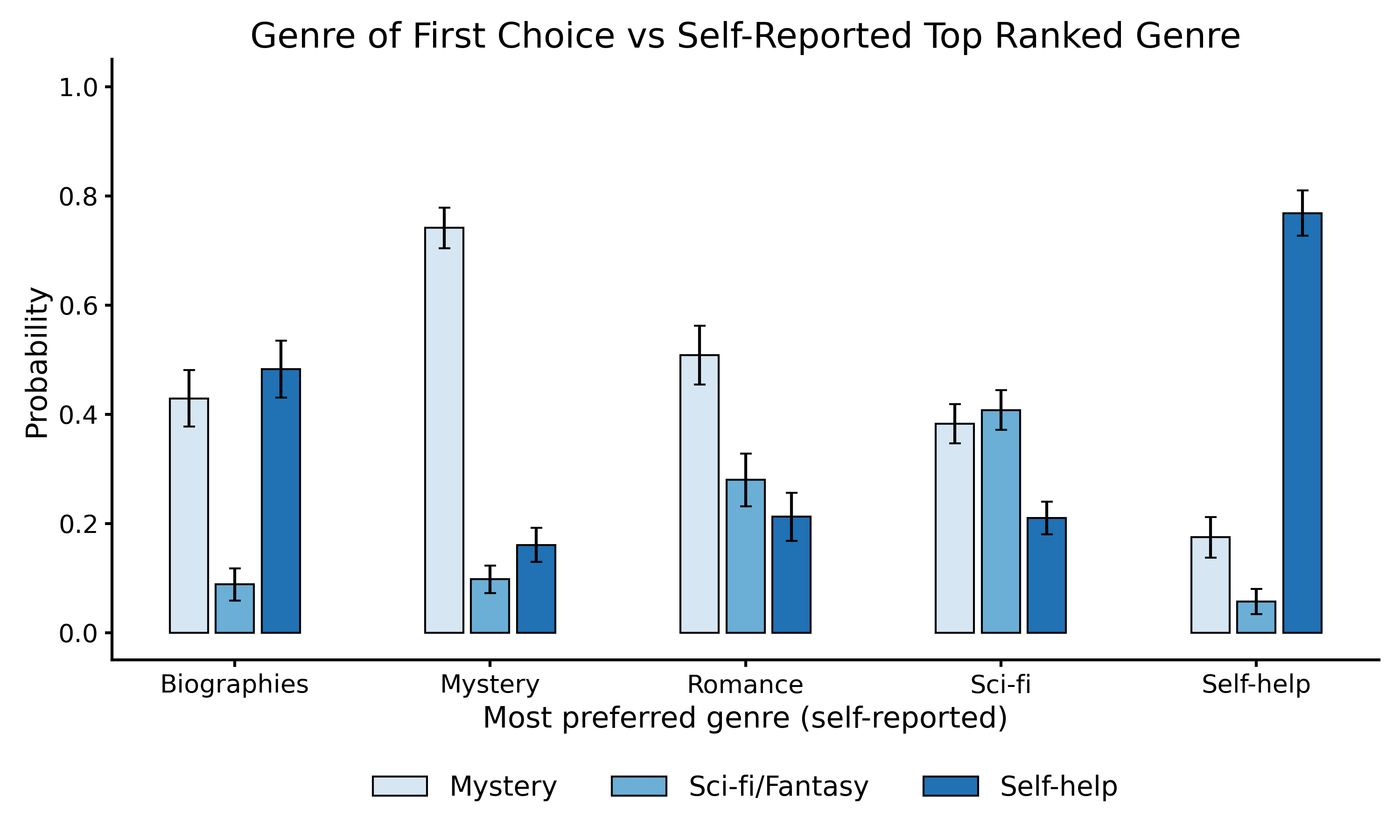}
\par\end{centering}
\caption{{\small\textbf{\protect\label{fig:genres_self_reported}Selected genres
and self-reported genre preferences in the experiment.}}}
\end{figure}
 
\begin{figure}[t]
\begin{centering}
\includegraphics[scale=0.42]{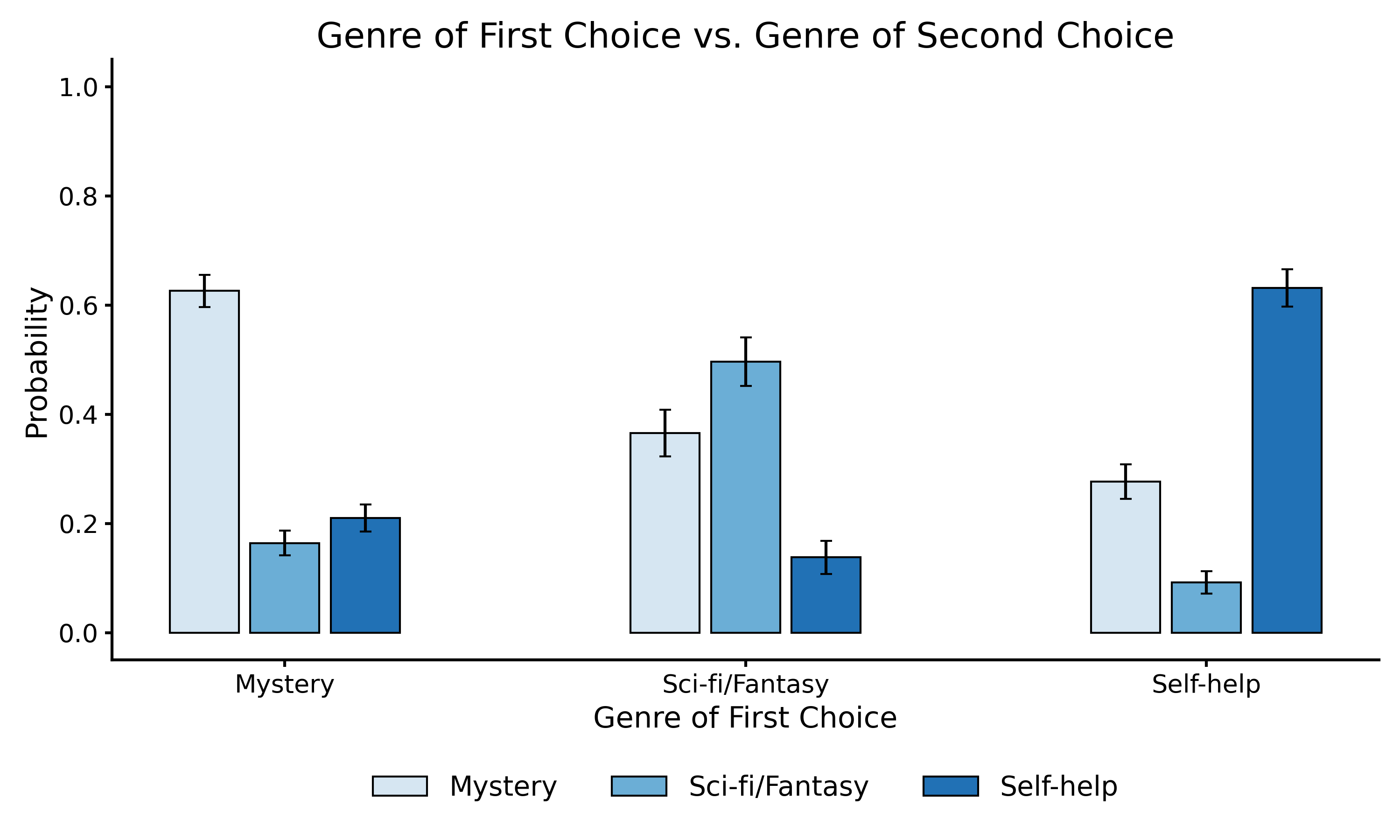}
\par\end{centering}
\caption{{\small\textbf{\protect\label{fig:genres_first_second_choices}Genres
of participants' first and second choices in the experiment.}}}
\end{figure}
\begin{figure}
\begin{centering}
\includegraphics[scale=0.7]{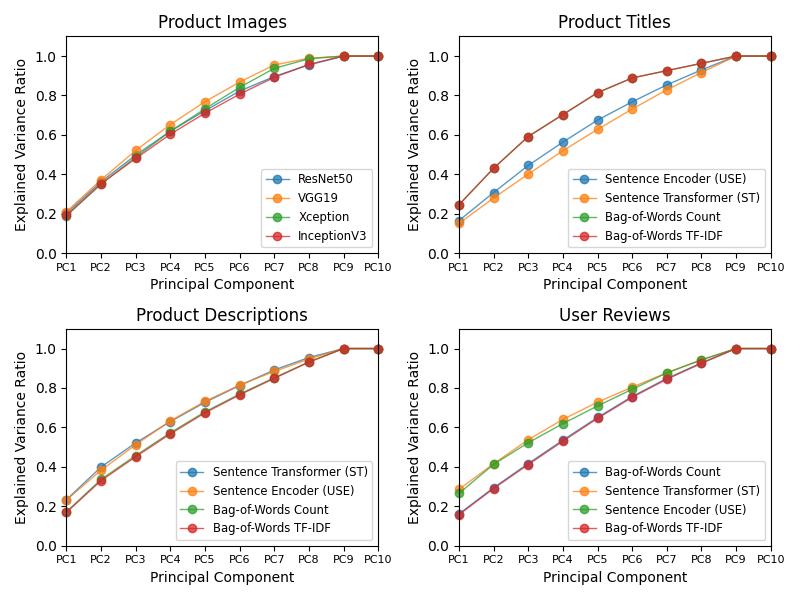}
\par\end{centering}
\caption{{\small\textbf{\protect\label{fig:pc_explained_variance}Share of
variance of embeddings explained by principal components.}}}
\end{figure}
\par\end{center}

\begin{center}
\begin{figure}
\centering{}\includegraphics[width=0.7\textwidth]{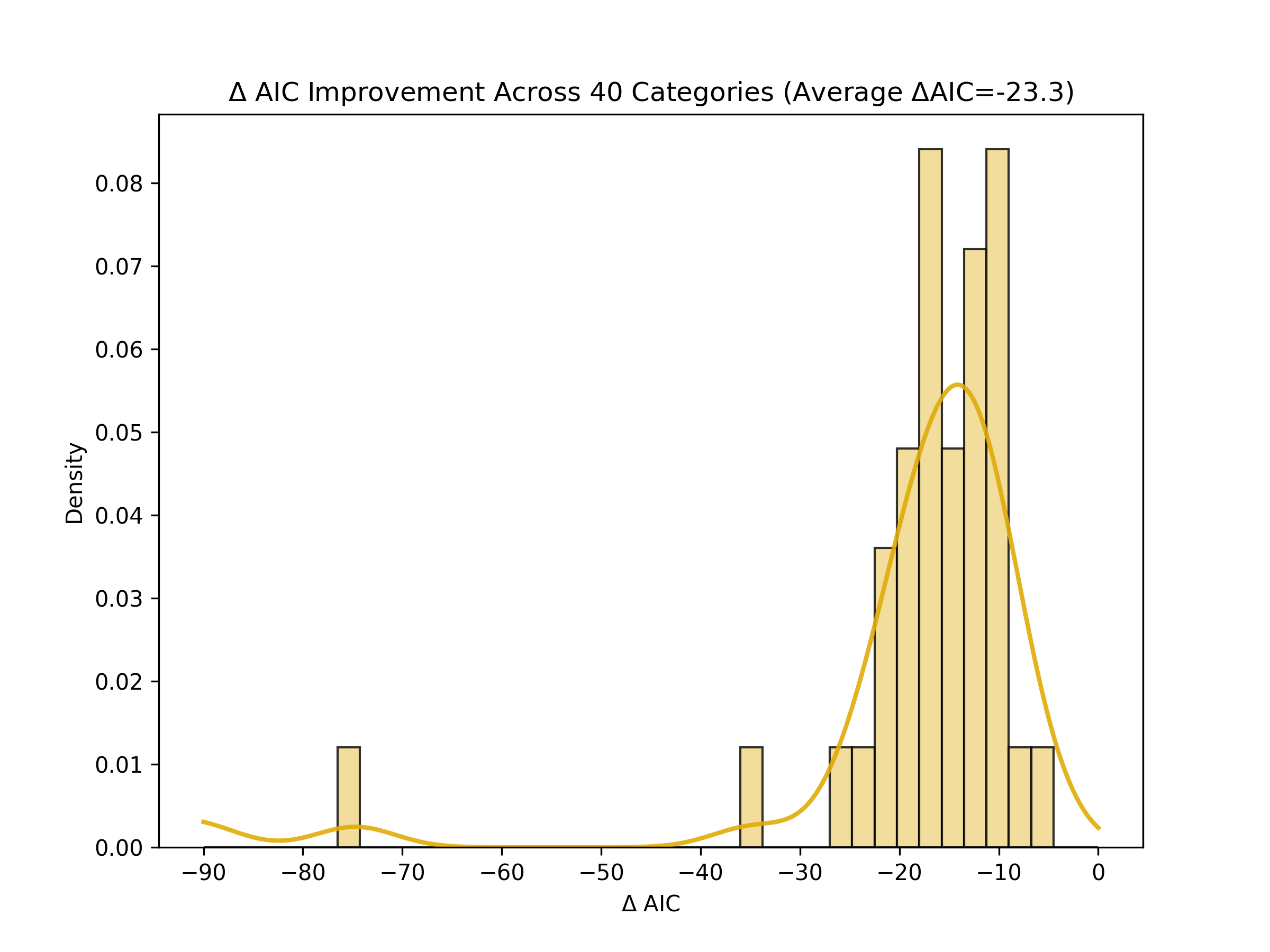}\caption{\textbf{\protect\label{fig:aic_improvements}}{\small\textbf{ Fit
improvements relative to plain logit across 40 product categories.}}{\small{}
$\Delta AIC$ in each category is the difference between the $AIC$
of the selected specification with unstructured data and that of the
plain logit model.}}
\end{figure}
\par\end{center}

\begin{center}
\begin{table}[h]
\begin{centering}
\resizebox{\textwidth}{!}{
\centering
\begingroup
\setlength{\tabcolsep}{12pt}
\begin{tabular}{l c c c c c}
\toprule
& \multicolumn{2}{c}{First Choices (Data)} & \multicolumn{3}{c}{Second Choices (Counterf.)} \\
\cmidrule(lr){2-3} \cmidrule(lr){4-6}
Model & $\log L$ & \textit{AIC} & \textit{RMSE} & \multicolumn{2}{c}{Rel. to Plain Logit} \\
 && & & $\Delta$ & \% \\
\midrule
\addlinespace
\multicolumn{6}{l}{\textbf{\textit{Panel A. Benchmark Model}}} \\
\addlinespace
Plain Logit & -20488.4 & 41006.7 & 0.091 &  &  \\
\addlinespace
\multicolumn{6}{l}{\textbf{\textit{Panel B. Mixed Logit Models with Principal Components}}} \\
\addlinespace
Images: InceptionV3 & -20475.2 & 40990.5 & 0.085 & -0.006 & -7.0\% \\
Images: ResNet50 & -20479.9 & 40997.8 & 0.083 & -0.009 & -9.4\% \\
Images: VGG19 & -20481.2 & 41000.4 & 0.087 & -0.005 & -5.0\% \\
Images: Xception & -20479.1 & 40996.3 & 0.089 & -0.002 & -2.1\% \\
Product Titles: Bag-of-Words Count & -20478.7 & 40995.5 & 0.084 & -0.007 & -7.9\% \\
Product Titles: Bag-of-Words TF-IDF & -20478.7 & 40995.5 & 0.084 & -0.007 & -7.9\% \\
Product Titles: Sentence Encoder (USE) & -20481.2 & 40998.3 & 0.087 & -0.004 & -4.3\% \\
Product Titles: Sentence Transformer (ST) & -20477.1 & 40992.3 & 0.085 & -0.007 & -7.3\% \\
Descriptions: Bag-of-Words Count & -20475.5 & 40988.9 & 0.085 & -0.006 & -6.9\% \\
Descriptions: Bag-of-Words TF-IDF & -20475.8 & 40989.6 & 0.085 & -0.006 & -6.5\% \\
Descriptions: Sentence Encoder (USE) & -20474.2 & 40986.4 & 0.076 & -0.015 & -17.0\% \\
Descriptions: Sentence Transformer (ST) & -20474.9 & 40987.8 & 0.075 & -0.016 & -18.1\% \\
Reviews: Bag-of-Words Count & -20479.6 & 40997.1 & 0.082 & -0.009 & -9.8\% \\
Reviews: Bag-of-Words TF-IDF & -20479.4 & 40996.9 & 0.082 & -0.009 & -10.1\% \\
\underline{Reviews: Sentence Encoder (USE)} & \underline{-20472.0} & \underline{40981.9} & \underline{0.070} & \underline{-0.021} & \underline{-23.0\%} \\
Reviews: Sentence Transformer (ST) & -20473.3 & 40984.6 & 0.073 & -0.019 & -20.4\% \\
\addlinespace
\multicolumn{6}{l}{\textbf{\textit{Panel C. Mixed Logit with Observed Attributes}}} \\
\addlinespace
Price & -20485.9 & 41003.9 & 0.091 & -0.000 & -0.0\% \\
Pages & -20484.9 & 41001.7 & 0.086 & -0.005 & -5.5\% \\
Year & -20484.7 & 41001.4 & 0.091 & 0.000 & 0.1\% \\
Genre & -20483.3 & 41000.7 & 0.084 & -0.007 & -7.7\% \\
Price \& Pages & -20478.9 & 40991.9 & 0.081 & -0.010 & -11.4\% \\
Price \& Year & -20484.1 & 41002.2 & 0.092 & 0.000 & 0.3\% \\
Price \& Genre & -20483.2 & 41002.5 & 0.086 & -0.005 & -5.2\% \\
\underline{Pages \& Year} & \underline{-20478.3} & \underline{40990.7} & \underline{0.081} & \underline{-0.011} & \underline{-11.7\%} \\
Pages \& Genre & -20481.6 & 40999.2 & 0.081 & -0.010 & -11.4\% \\
Year \& Genre & -20483.3 & 41002.7 & 0.084 & -0.007 & -7.7\% \\
Price, Pages, \& Year & -20478.3 & 40992.7 & 0.081 & -0.011 & -11.7\% \\
Price, Pages, \& Genre & -20478.9 & 40995.9 & 0.081 & -0.010 & -11.4\% \\
Price, Year, \& Genre & -20482.2 & 41002.4 & 0.084 & -0.007 & -7.4\% \\
Pages, Year, \& Genre & -20478.3 & 40994.7 & 0.081 & -0.011 & -11.7\% \\
Price, Pages, Year, \& Genre (All Attr.) & -20476.9 & 40993.9 & 0.078 & -0.013 & -14.1\% \\
\bottomrule
\end{tabular}
\endgroup
}
\par\end{centering}
\caption{{\small\textbf{\protect\label{tab:validation_numbers}Model validation
results. }}{\small The table shows in-sample fit on first-choice data
and counterfactual performance on second-choice data for all specifications
considered in Figure }{\small\textbf{\ref{fig:model_comparison_rmse}}}{\small{}
(see Section \ref{subsec:model_specifications} for detailed description
of these models).}}
\end{table}
\par\end{center}

\begin{center}
\begin{table}
\begin{centering}
\resizebox{0.95\textwidth}{!}{\footnotesize
\begin{tabular}{l|*{8}{>{\centering\arraybackslash}p{1.3cm}}}
\toprule
  & Fire 7 & Fire HD & Fire HD & Fire 7 & iPad & Fire HD & Dragon & iPad \\ 
  &  & 8 & 10 & Kids & 10.2 & 8 Kids & Touch & 9.7 \\
\midrule
Fire 7 & 0.000 & 0.330 & 0.381 & 0.301 & 0.312 & 0.283 & 0.279 & 0.286 \\
Fire HD 8 & 0.225 & 0.000 & 0.229 & 0.175 & 0.178 & 0.165 & 0.162 & 0.166 \\
Fire HD 10 & 0.386 & 0.340 & 0.000 & 0.314 & 0.320 & 0.297 & 0.291 & 0.295 \\
Fire 7 Kids & 0.123 & 0.105 & 0.128 & 0.000 & 0.100 & 0.091 & 0.089 & 0.091 \\
iPad 10.2 & 0.153 & 0.129 & 0.148 & 0.122 & 0.000 & 0.113 & 0.112 & 0.114 \\
Fire HD 8 Kids & 0.043 & 0.038 & 0.049 & 0.035 & 0.035 & 0.000 & 0.032 & 0.033 \\
Dragon Touch & 0.022 & 0.018 & 0.021 & 0.017 & 0.017 & 0.016 & 0.000 & 0.016 \\
iPad 9.7 & 0.048 & 0.041 & 0.044 & 0.037 & 0.037 & 0.035 & 0.034 & 0.000 \\
\bottomrule
\end{tabular}}
\par\end{centering}
\caption{{\small\textbf{\protect\label{tab:diversions_logit}Estimated diversion
ratios for tablets (plain logit). }}{\small Each cell reports the probability
of choosing the row product $j$ when the first choice, column product
$k$, is removed from the choice set.}}
\end{table}
\par\end{center}

\begin{center}
\begin{table}
\begin{centering}
\resizebox{0.95\textwidth}{!}{\footnotesize
\begin{tabular}{l|*{8}{>{\centering\arraybackslash}p{1.3cm}}}
\toprule
  & Fire 7 & Fire HD & Fire HD & Fire 7 & iPad & Fire HD & Dragon & iPad \\ 
  &  & 8 & 10 & Kids & 10.2 & 8 Kids & Touch & 9.7 \\
\midrule
Fire 7 & 0.000 & 0.939 & 0.451 & 0.144 & 0.328 & 0.019 & 0.347 & 0.073 \\
Fire HD 8 & 0.592 & 0.000 & 0.027 & 0.023 & 0.068 & 0.002 & 0.267 & 0.008 \\
Fire HD 10 & 0.247 & 0.023 & 0.000 & 0.167 & 0.213 & 0.328 & 0.299 & 0.332 \\
Fire 7 Kids & 0.035 & 0.006 & 0.090 & 0.000 & 0.055 & 0.624 & 0.000 & 0.016 \\
iPad 10.2 & 0.100 & 0.025 & 0.147 & 0.072 & 0.000 & 0.018 & 0.058 & 0.558 \\
Fire HD 8 Kids & 0.004 & 0.000 & 0.174 & 0.582 & 0.009 & 0.000 & 0.000 & 0.011 \\
Dragon Touch & 0.003 & 0.004 & 0.006 & 0.000 & 0.001 & 0.000 & 0.000 & 0.002 \\
iPad 9.7 & 0.018 & 0.002 & 0.105 & 0.012 & 0.326 & 0.009 & 0.029 & 0.000 \\
\bottomrule
\end{tabular}}
\par\end{centering}
\caption{{\small\textbf{\protect\label{tab:diversions_attributes}Estimated
diversion ratios for tablets (mixed logit with observed attributes).
}}{\small Each cell reports the probability of choosing the row product
$j$ when the first choice, column product $k$, is removed from the
choice set.}}
\end{table}
\par\end{center}

\begin{center}
\begin{table}
\begin{centering}
\resizebox{0.95\textwidth}{!}{\footnotesize
\begin{tabular}{l|*{8}{>{\centering\arraybackslash}p{1.3cm}}}
\toprule
  & Fire 7 & Fire HD & Fire HD & Fire 7 & iPad & Fire HD & Dragon & iPad \\ 
  &  & 8 & 10 & Kids & 10.2 & 8 Kids & Touch & 9.7 \\
\midrule
Fire 7 & 0.000 & 0.590 & 0.162 & 0.362 & 0.143 & 0.033 & 0.024 & 0.080 \\
Fire HD 8 & 0.753 & 0.000 & 0.429 & 0.336 & 0.200 & 0.193 & 0.108 & 0.116 \\
Fire HD 10 & 0.058 & 0.144 & 0.000 & 0.079 & 0.142 & 0.297 & 0.022 & 0.077 \\
Fire 7 Kids & 0.127 & 0.131 & 0.080 & 0.000 & 0.012 & 0.440 & 0.293 & 0.007 \\
iPad 10.2 & 0.049 & 0.079 & 0.141 & 0.010 & 0.000 & 0.015 & 0.310 & 0.715 \\
Fire HD 8 Kids & 0.003 & 0.034 & 0.151 & 0.200 & 0.004 & 0.000 & 0.188 & 0.001 \\
Dragon Touch & 0.000 & 0.001 & 0.000 & 0.011 & 0.013 & 0.019 & 0.000 & 0.003 \\
iPad 9.7 & 0.011 & 0.020 & 0.036 & 0.002 & 0.485 & 0.003 & 0.055 & 0.000 \\
\bottomrule
\end{tabular}}
\par\end{centering}
\caption{{\small\textbf{\protect\label{tab:diversions_unstructured}Estimated
diversion ratios for tablets (selected model: }}{\small\textbf{\textit{Descriptions
ST}}}{\small\textbf{). }}{\small Each cell reports the probability
of choosing the row product $j$ when the first choice, column product
$k$, is removed from the choice set.}}
\end{table}
\par\end{center}

\begin{center}
\begin{table}
\begin{centering}
{\small{\footnotesize
\begin{tabular}{p{5cm}>{\centering\arraybackslash}p{2cm}>{\centering\arraybackslash}p{2cm}>{\centering\arraybackslash}p{1.5cm}>{\centering\arraybackslash}p{3cm}}
\toprule
Category & Data Type & Model Type & $\Delta$ AIC & $\Delta$ Diversion to Closest Substitute \\
\midrule
1. Clothing Active & Images & VGG16 & -25.1 & 11.5\% \\
2. Clothing Shirts & Reviews & USE & -11.9 & 12.5\% \\
3. Clothing Underwear & Titles & TFIDF & -16.3 & 45.5\% \\
4. Clothing Sleep & Images & Inceptionv3 & -13.1 & 32.5\% \\
5. Clothing Tops \& Blouses & Titles & USE & -12.3 & 30.4\% \\
6. Electronics Cables & Reviews & TFIDF & -16.7 & 20.1\% \\
7. Electronics Accessories & Descriptions & COUNT & -19.6 & 19.0\% \\
8. Electronics Keyboards & Descriptions & ST & -19.5 & 20.0\% \\
9. Electronics Memory Cards & Images & VGG16 & -96.8 & 33.4\% \\
10. Electronics Tablets & Descriptions & ST & -111.5 & 18.2\% \\
11. Electronics Monitors & Images & Resnet50 & -22.4 & 19.9\% \\
12. Electronics Headphones & Images & VGG19 & -10.7 & 6.1\% \\
13. Electronics Media Players & Images & VGG19 & -90.2 & 23.4\% \\
14. Groceries Water & Titles & USE & -19.1 & 19.0\% \\
15. Groceries Coffee & Descriptions & TFIDF & -9.5 & 31.8\% \\
16. Groceries Tea & Images & Xception & -17.1 & 37.4\% \\
17. Groceries Chips & Descriptions & ST & -5.6 & 19.2\% \\
18. Household Aromatherapy & Descriptions & COUNT & -11.2 & 17.6\% \\
19. Household Batteries & Images & VGG16 & -10.7 & -0.6\% \\
20. Household Trash Bags & Descriptions & TFIDF & -16.2 & 40.1\% \\
21. Household Paper Towels & Titles & TFIDF & -16.9 & 33.1\% \\
22. Home Sheets \& Pillowcases & Images & Resnet50 & -21.3 & 6.8\% \\
23. Bedroom Beds & Descriptions & USE & -11.2 & 26.0\% \\
24. Bedroom Mattresses & Reviews & ST & -12.9 & 23.5\% \\
25. Kitchen Food Storage & Images & VGG19 & -20.5 & 28.9\% \\
26. Office Folders & Descriptions & ST & -17.4 & 31.9\% \\
27. Office Paper & Reviews & ST & -13.5 & 21.0\% \\
28. Office Markers & Images & Inceptionv3 & -13.6 & 19.8\% \\
29. Office Pens & Images & VGG19 & -13.5 & 31.6\% \\
30. Office Printer Supplies & Titles & USE & -15.1 & 34.1\% \\
31. Pet Cat Food & Descriptions & COUNT & -11.1 & 36.2\% \\
32. Pet Cat Litter & Titles & ST & -12.1 & 34.1\% \\
33. Pet Cat Snacks & Reviews & COUNT & -16.9 & 29.1\% \\
34. Pet Dog Food & Images & Inceptionv3 & -11.7 & 24.5\% \\
35. Pet Dog Treats & Titles & USE & -18.2 & 38.0\% \\
36. Game Consoles Nintendo & Images & Resnet50 & -6.8 & 11.9\% \\
37. Video Games Nintendo & Descriptions & ST & -23.5 & 13.3\% \\
38. Video Games PC & Titles & COUNT & -9.5 & 20.7\% \\
39. Video Games PS4 & Descriptions & ST & -34.7 & 32.2\% \\
40. Video Games Xbox & Images & Xception & -74.7 & 28.9\% \\
\bottomrule
\hline
Averaged &  &  & -23.3 & 24.6\% \\
\end{tabular}
}}
\par\end{centering}
\caption{{\small\textbf{\protect\label{tab:category_level_results}Estimation
results across 40 categories in Comscore data. }}{\small For each category,
the table shows the selected model and data type yielding the lowest
$AIC$ and the $AIC$ improvement relative to plain logit. The last
column shows the increase in the predicted diversion ratios to the
closest substitutes, $\max_{k}\ensuremath{\hat{s}_{j\rightarrow k}}$,
averaged across products $j$, relative to plain logit (in percentage
points).}}
\end{table}
\par\end{center}

\begin{center}
\par\end{center}
\end{document}